\documentclass[usenatbib]{mnras}
\usepackage{newtxtext,newtxmath}
\usepackage[T1]{fontenc}
\usepackage{ae,aecompl}

\usepackage{upgreek}
\usepackage{orcidlink}

\usepackage{graphicx}	
\usepackage{amsmath}	
\usepackage{multirow}
\usepackage{hyperref}
\usepackage{xspace}
\usepackage{bm}

\newcommand{\rhotot}{\rho_{\mathrm{total}}}
\newcommand{\rhototdot}{\dot\rho_{\mathrm{total}}}
\newcommand{\rhoa}{\rho_{\mathrm{a}}}
\newcommand{\rhob}{\rho_{\mathrm{b}}}
\newcommand{\rhoc}{\rho_{\mathrm{c}}}
\newcommand{\rhoi}{\rho_{i}}
\newcommand{\taua}{\tau_{\mathrm{a}}}
\newcommand{\taub}{\tau_{\mathrm{b}}}
\newcommand{\tauc}{\tau_{\mathrm{c}}}
\newcommand{\taui}{\tau_{i}}

\newcommand{\fV}{V} 
\newcommand{\nelec}{n_{\rm e}}
\newcommand{\meas}{}
\newcommand{\autoreffigs}[1]{\hyperref[#1]{Figures~\ref*{#1}}}
\newcommand{\autorefb}[1]{\hyperref[#1]{Equation~(\ref*{#1})}}
\newcommand{\autorefp}[1]{\hyperref[#1]{(Equation~\ref*{#1})}}

\newcommand{\revone}[1]{#1}

\newcommand{\arctic}{\texttt{ArCTIc}\xspace}
\newcommand{\acswfc}{\textit{ACS}/\textit{WFC}\xspace}
\newcommand{\gresim}{\ {\raise-.5ex\hbox{$\buildrel>\over\sim$}}\ }
\newcommand{\lessim}{\ {\raise-.5ex\hbox{$\buildrel<\over\sim$}}\ }
\def \degC{$^\circ\kern-0.06em\rm{C}$} 

\title[Radiation damage to HST, corrected with ArCTIc]{Radiation damage to the Hubble Space Telescope during two Solar cycles,\\ and 
correction of Charge Transfer Inefficiency using ArCTIc
} 
\author[R.\ Massey et al.]{Richard Massey$^{1}$\orcidlink{0000-0002-6085-3780},
Jacob A.\ Kegerreis$^{2,3,1}$\orcidlink{0000-0001-5383-236X},
Juan Paolo Lorenzo Gerardo Barrios$^{4,1}$\orcidlink{0000-0002-5605-0029},
\newauthor 
James W.\ Nightingale$^{5,1}$\orcidlink{0000-0002-8987-7401},
Richard G.\ Hayes$^{1}$,
David Lagattuta$^{1}$\orcidlink{0000-0002-7633-2883}, 
Zane D.\ Lentz$^{1}$\orcidlink{0000-0003-4428-7843}, 
\newauthor
Gavin Leroy$^{1}$\orcidlink{0009-0004-2523-4425},
Jesper Skottfelt$^{6}$\orcidlink{0000-0003-1310-8283},
\revone{Felix Vecchi$^{7}$\orcidlink{0009-0004-7808-1979}}
and
Maximilian von Wietersheim-Kramsta$^{1}$\orcidlink{0000-0003-4986-5091}\\
$^{1}$Institute for Computational Cosmology, Durham University, South Road, Durham DH1 3LE, UK\\
$^{2}$Department of Earth Science and Engineering, Imperial College London, London SW7 2BP, UK\\
$^{3}$SETI Institute, 339 Bernardo Avenue, Suite 200, Mountain View, CA 94043, USA\\
$^{4}$Cavendish Laboratory, University of Cambridge, JJ Thomson Avenue, Cambridge CB3 0HE, UK\\
$^{5}$Physics Department, Newcastle University, Newcastle upon Tyne NE1 7RU, UK\\
$^{6}$Centre for Electronic Imaging, Department of Physical Sciences, The Open University, Walton Hall, Milton Keynes MK7 6AA, UK\\
$^{7}$Laboratoire d’Astrophysique, EPFL, Observatoire de Sauverny, 1290 Versoix, Switzerland\vspace{-6mm}
}
\date{23 November 2025\vspace{-3mm}
}
\pubyear{2025}

\begin{document}
\label{firstpage}
\pagerange{\pageref{firstpage}--\pageref{lastpage}}
\maketitle

\begin{abstract}
From 2002 to 2025, the Hubble Space Telescope’s Advanced Camera for Surveys has suffered in the harsh radiation environment above the protection of the Earth’s atmosphere.
We track the degradation of its image quality, as Solar protons and galactic cosmic rays have damaged its photosensitive charge-coupled device (CCD) imaging sensors.
The rate of damage in low Earth orbit is modulated by $18.5^{+4.5}_{-0.5}$~per cent during an 11 year Solar cycle, peaking $430^{+11}_{-5}$~days after Solar minimum as recorded in the number of sunspots.
The type of damage is consistent with defects in the silicon lattice that have all stabilised into one of three configurations.
We also present the open-source Algorithm for Charge Transfer Inefficiency correction (ArCTIc) v7, available from \href{https://github.com/jkeger/arctic}{https://github.com/jkeger/arctic}.
This models the (instantaneous or gradual) capture of photoelectrons into lattice defects, and their release after (a discrete set or continuum of) characteristic time delays, 
which creates spurious trailing in an image.
Calibrated using the trailing of hot pixels, and applied during post-processing of astronomical images, ArCTIc can correct 99.5\% of Charge Transfer Inefficiency trailing averaged over the camera's lifetime, and 99.9\% of trailing in the worst-affected recent data.\\
\end{abstract}

\begin{keywords}\vspace{-8mm}
space vehicles: instruments -- 
instrumentation: detectors -- 
Sun: particle emission -- 
sunspots --
software: data analysis 
\end{keywords}

\vspace{-10mm}
\section{Introduction}

\begin{figure*}
	\centering
	\includegraphics[
    width=\textwidth, trim={37mm 8mm 24mm 25mm}, clip]{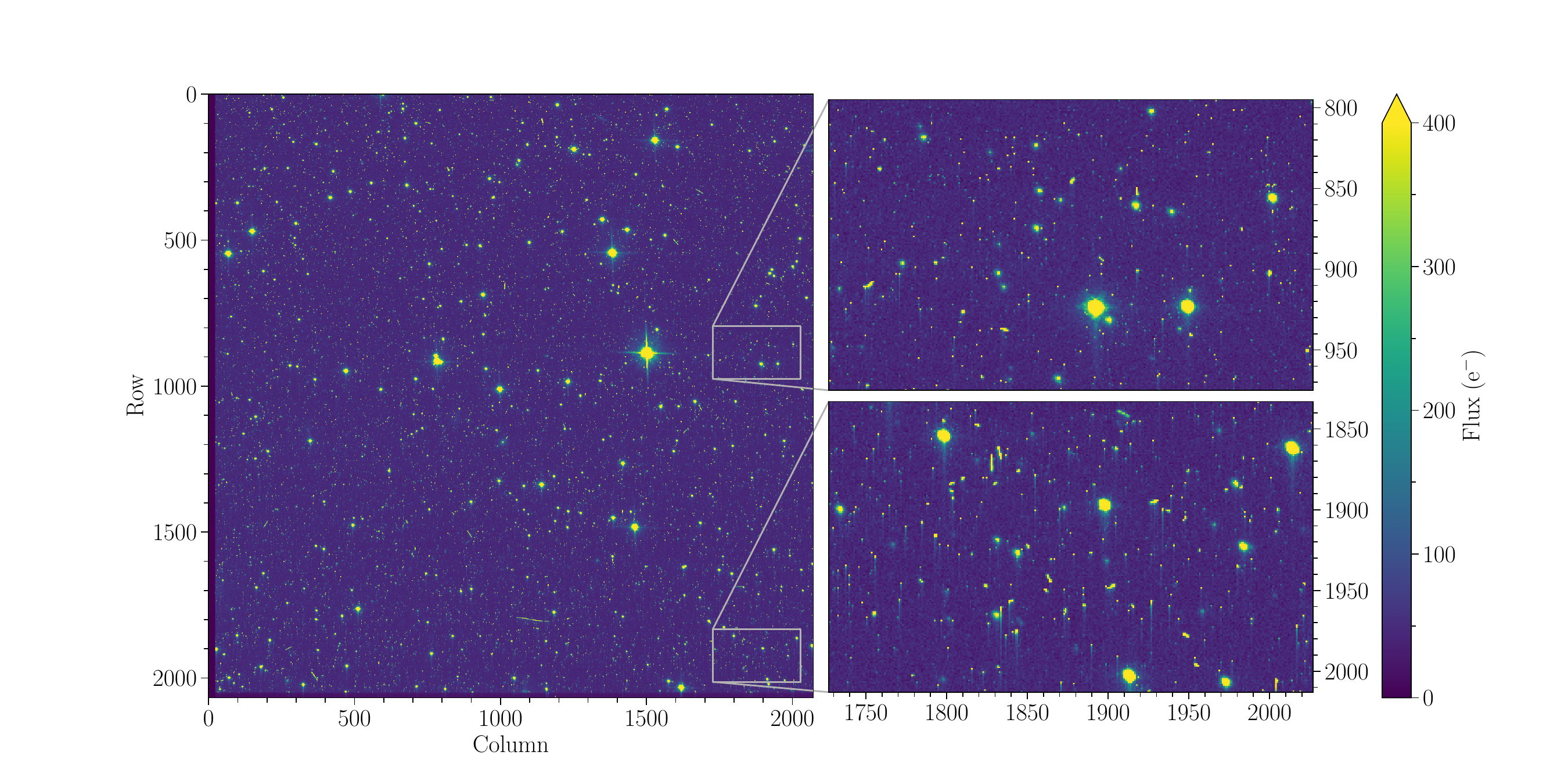}
  \caption{
    An example {\sl HST} \acswfc image to illustrate our typical data and 
    the effects of CTI (quadrant D of an image taken on 2021 January 3, approximately 19~years after {\sl ACS} was launched).
    The readout register is at the top of this image, above row 0.
    Electrons from the first rows of pixels undergo few transfers,
    encounter not many charge traps, and thus do not suffer much CTI trailing.
    Electrons from pixels further down the page undergo many transfers,
    so significant charge can be captured and later released by traps.
    This effect creates visible CTI trails in the upper zoom region
    and even worse trails in the lower zoom.
    The edge regions of low flux at the left and bottom of the image
    are the serial prescan and parallel overscan. 
    That these virtual pixels should contain zero electrons in an image 
    is a convenient validation and verification test for successful CTI correction.
    \label{fig:example_image_zooms}}
\end{figure*}

Charge-coupled device (CCD) imaging detectors work by converting photons into electrons, then shepherding groups (pixels) of those electrons to the edge of the device for counting.
Electrons can be transferred very efficiently through a pristine silicon wafer.
However, detectors kept in high radiation environments (e.g.\ space, particle or medical devices) accumulate lattice defects that impede the free movement of electrons \citep{Holland1990}.
Electrons get temporarily trapped in the defects; they are released after a characteristic delay \citep{ShockleyRead1952,Hall1951}, but can thus become separated from the other electrons in their original pixel.
The released electrons re-appear as spurious trails behind every object in an image \citep{Janesick1987,Jerram2020,Ali2022}, 
as illustrated in \autoref{fig:example_image_zooms}.
Different `species' of traps (different configurations of the damaged silicon lattice) capture electrons for different lengths of time, resulting in a superposition of various trail profiles.

This charge transfer inefficiency (CTI) is particularly troublesome because the number of trailed electrons is a nonlinear function of the brightness, size, and shape of an image feature, and of the previous illumination history of nearby pixels \citep{Rhodes2007}. 
Because the effect is not linear, it cannot be described by a convolution. 
To first order, this can be roughly understood by considering that there are only a finite number of charge traps, so the {\it fraction} of electrons trailed from faint objects will be greater than that from bright ones. 

Trailing can be corrected during data postprocessing.
Since CCD readout is (almost) the last process to happen during data acquisition, it should be corrected during the first stage of data analysis.
The process of {\it adding} CTI trails can be reproduced in software that models the flow of electrons past the charge traps. 
\citet{Bristow2003} pioneered an iterative method to {\it remove} trailing in one of {\sl HST}'s first-generation cameras: by repeatedly adding trails to a model of the true data, and adjusting the model. 
After the {\it Advanced Camera for Surveys} ({\it ACS}) was installed on 2002 March 1, \cite{Massey+2010} used the method to correct 97\% of the trailing in {\it ACS}/{\it Wide Field Channel} ({\it WFC}) data up to 2006, and the approach was incorporated into the STScI {\tt CALACS} software by \cite{Anderson2010}.
As radiation damage accumulated, the trailing increased and became easier to measure.
\cite{Massey+2014} achieved 98\% correction, and STScI continue to update the algorithm \citep{Chiaberge2022,Anderson2024,Ryon2024}.

Although most of the trailing can now be removed, the remainder still limits high precision science with {\sl HST}.
There are many examples where measurements of object photometry, astrometry or morphology are required to better than 1\% accuracy \citep[e.g.][]{Ghez2008,Cropper2013,Ahmed2020,Mitra2021,Zhao2021,Kelman2023,Soto2023,Astier2023,Stark2024,2025hst..prop17925B,plato2025} -- and {\sl HST}'s detectors continue to degrade.
To investigate how accurate a correction is feasible after many years of calibration data, this paper attempts to build a new model for {\sl HST} \acswfc\ at better than 1\% precision, and to introduce a flexible tool that can be used to correct CTI for a wide variety of applications.

The rate of damage to {\sl HST} is also a proxy for the flux of high-energy radiation in low Earth orbit (LEO). Exploiting uniform in-orbit measurements from a single, highly sensitive camera, we shall track how fluctuations in the LEO radiation environment depend upon Solar activity spanning more than two full Solar cycles from 2002 to 2025.

This paper is organised as follows. We describe the \arctic algorithm in \S\ref{sec:arctic}. We calibrate its parameters for {\sl HST}, measuring the rate of radiation damage, in \S\ref{sec:calibration}. We correct CTI and quantify \arctic's performance in \S\ref{sec:results}. We discuss limitations and ideas for future work in \S\ref{sec:discussion}, then conclude in \S\ref{sec:conclusions}.

\section{ArCTIc: Algorithm for CTI correction} \label{sec:arctic}

We aim to reproduce in (fast) software the spurious trailing of charge as it traverses a CCD during readout. This requires: shuffling packets of electrons through many pixels; and monitoring the occupancy of charge traps within those pixels, as they capture or release electrons. 

Version 7 of \arctic, written in parallelised C++ with a python wrapper, is open-source and available with examples, documentation, and unit tests at \href{https://github.com/jkeger/arctic}{github.com/jkeger/arctic}. Compared to previous versions, it provides both a significant speed improvement and new model features. 
Typical runtime is $\sim$1\,second per {\sl ACS/WFC} CCD quadrant.
The code incorporates a variety of configurable models and options for properties of the CCD, charge traps, and readout electronics (ROE).
Although we shall describe algorithms in terms of the movement of electrons, as appropriate for n-type CCDs, the algorithm works equally well for the movement of holes in p-type CCDs \citep[e.g.][]{Tarle2003}. The code can also operate in both directions of parallel (`$y$') and serial (`$x$') CCD readout, even though the particular species of defects in {\sl ACS/WFC}, as well as its operating temperature and clock speed, happen to make parallel CTI far more significant than serial CTI \citep{Rivers2023,Ryon2024}.

\subsection{ArCTIc model of a CCD} \label{sec:arctic:ccd}

\arctic assumes a `volume-driven' model of a CCD \citep{Short2013}, in which a packet of $\nelec$ electrons is assumed to occupy some fraction, $0\leqslant\fV\leqslant1$, of the volume of a pixel (or phase within a pixel). The function $\fV(\nelec)$ must be monotonically non-decreasing and, by default, we use
\begin{equation}
  \fV(\nelec) = 
  \left(\dfrac{\nelec - d}{w - d}\right)^\beta 
  \,,
  \label{eqn:cloud_volume}
\end{equation}
clamped between 0 and 1,
where $d$ is the depth of the supplementary buried channel or `notch' \revone{in units of electrons, $w$ the full well capacity in units of electrons,} and $\beta$ the well fill power.

Note that clamping $\fV$\,$\geqslant$\,$0$ has a consequence for faint (e.g.\ bias) images. Although the {\it true} number of electrons in a pixel must physically be non-negative, read noise can suggest pixel values of $\nelec-d<0$, even if $d=0$. Positive noise spikes above $d$ get trailed, but positive spikes below $d$ do not. All negative noise spikes are filled by trails, and the total number of electrons is conserved, but this asymmetry in the presence of noise is a choice (discussed further in \S\ref{sec:discussion}).

\subsection{ArCTIc model of charge traps} \label{sec:arctic:traps}

Encouraged by high precision measurements from CCDs irradiated on the ground \citep{C3TM,Bilgi2019}, we adopt a physically-motivated model of silicon defects, in which each trap delays electrons probabilistically but with a characteristic half-life time \citep[as opposed to a pixellated trail profile, like][]{Anderson2010}.  
In this model, CCD defects belong to discrete species~$i$,  corresponding to topological configurations of silicon atoms, or impurities of different elements \citep{Skottfelt2022}.

We assume that traps are distributed with spatially uniform\footnote{Vertical gradients of traps are easily accommodated, by adjusting the functional form of \autorefb{eqn:cloud_volume}. This includes  `surface' traps that are usually associated with front-side electronics or manufacturing defects, and capture electrons only from almost-saturated pixels. That \autorefb{eqn:exponential_model}'s linear dependence upon $y$ fits well (see \S\ref{sec:trail_fits}) demonstrates that there must be negligible large-scale gradient of traps in the $y$-direction.} effective density per pixel, $\rho_i$. 
In a real device, Poisson statistics \citep{Ogaz+2013}, and catastrophic events \citep{Guzman2024} inevitably mean that some pixels or columns contain more defects than others. 
Nonetheless, the assumption of uniform density remains acceptable because, during readout, most packets of electrons traverse thousands of pixels whose effects average out; and many science requirements for CTI correction also concern statistical ensembles \citep[e.g.][]{Cropper2013}. Furthermore, unless the true location of every trap is known, assuming uniform density avoids a source of discretisation noise when placing model traps.
To eliminate a second source of discretisation noise, \arctic also models `fractional traps' that capture and release {\it fractions} of electrons. This marginalises over the quantum mechanical probability density function for the stochastic processes of capture and release.
The output image from the current algorithm is identical to the mean output of a previous version of \arctic that randomly distributed traps, each of which captured integer numbers of electrons \citep{Rhodes+2010}. 
But when every pixel is identical, runtime can be reduced by orders of magnitude.

During each step of readout, \arctic\ monitors the occupancy fraction $0\leqslant f\leqslant 1$ of charge traps.
To further reduce runtime, traps are grouped into batches that must share the same occupancy fraction. These groups are between `watermarks' analogous to marks on a gauge left by a high tide or coastal wave. Whenever a packet of electrons passes through a pixel, it fills all traps up to some `height' (the fractional volume of the pixel $\fV$ occupied by free electrons according to \autoref{eqn:cloud_volume}). Traps above this watermark are unaffected, so higher watermarks from previous waves of electrons remain (progressively drying, as captured electrons are gradually released).

Algorithmically, the watermarks are tracked using one shared array to store their height, plus one array per trap species to store the current occupancy of traps $f$ between that watermark and the watermark below.
The size of these arrays can change.
Runtime is significantly improved by storing watermarks in volume order, and by initialising large arrays containing sufficient elements that, if need be, a new watermark can be created after every pixel-to-pixel transfer. Rather than dynamically allocating memory to change the size of the arrays, \arctic monitors which array elements are active, and shuffles those around as needed.

\subsubsection{Instantaneous capture and homogeneous release} \label{sec:traps:instant}

Unoccupied (or partially unoccupied) charge traps inside the volume occupied by a packet of electrons can capture an electron. The simplest algorithm assumes that capture is much faster than both the dwell time that electrons spend in the pixel and the trap release time. Thus, if a packet of $\nelec$ electrons moves into a hitherto empty pixel, 
\begin{equation}
\label{eqn:instant_capture_traps_captured}
  n_{\rm capture} (\nelec) = \sum_i^{\rm species} \rho_{i} ~ \fV(\nelec)
\end{equation} 
electrons are instantaneously captured, 
where the summation runs over trap species, each of which has effective density per pixel $\rho_{i}>0$. 
A new watermark is created at the level of the cloud's effective volume $\fV(\nelec)$. All traps below it become fully filled ($f$=1), and any pre-existing, partially-filled watermarks below it are discarded.
The number of captured electrons, $n_{\rm capture}$, are removed from the packet of free electrons.

Charge traps above the current watermark release electrons. 
If the initial occupancy of a trap is $f(t_{\rm 0})$, after additional time $t_{\rm dwell}$, it becomes 
\begin{equation}
  f(t_{\rm 0}+t_{\rm dwell}) = f(t_{\rm 0}) \exp\left[-\,\dfrac{t_{\rm dwell}}{\tau^{\rm rel}_i}\right] \,,
  \label{eqn:instant_capture_occupancy_change}
\end{equation}
where the characteristic release time of each trap species, $\tau^{\rm rel}_i$, depends on the physical configuration of the trap and the temperature of the CCD. 
Values of $\tau^{\rm rel}_i$ are conveniently stored in units of $t_{\rm dwell}$, so they become the characteristic length of a trail in pixels. 
When releasing charge, the occupancy levels change but the watermark levels do not. The number of electrons released back into the packet is
\begin{equation}
  n_{\rm release} = \sum_j^{\rm watermarks} \sum_i \, \rho_{i} \big( \fV_{j+1} - \fV_{j} \big)\,
  f(t_{\rm 0}+t_{\rm dwell})\,.
  \label{eqn:instant_capture_traps_released}
\end{equation}

The order of capture and release operations is important.
Because in this model release is slow, we first release charge from any traps above the current watermark. 
This may slightly raise the watermark.
We then capture electrons in traps up to the updated watermark, because capture in this model is instantaneous, and the steady state of all traps surrounded by electrons should be filled.
In rare situations (usually with very low $\beta$) where insufficient electrons are available to fill traps below the current watermark, the occupancy fractions are increased part-way to full, all by the same fractional change.

The trail profile produced by this algorithm behind a sharp image feature, is roughly a sum of decaying exponentials. It is more complex than that, because some
trailed electrons are recaptured then trailed again \citep[see Figure~1 of][]{Massey2010}. Nonetheless, it will be useful (in \S\ref{sec:remove_cti} and \S\ref{sec:trail_fits:otf_validation}) to have a closed-form analytic approximation to the trail behind a delta-function image, such as a warm pixel. This is the Green's function of CTI, if you will:
\begin{eqnarray}
  \nelec^\mathrm{trail}(\Delta y) \approx \left[
    \left(\dfrac{\nelec^\mathrm{wp} - d}{w - d}\right)^\beta
    - \left(\dfrac{\nelec^{\rm bg} - d}{w - d}\right)^\beta \right]
    y ~~~~~~~~~~~~~~~~~~~~~~~~ \nonumber \\
 \times~\sum_i^{\mathrm{species}} \rhoi
      \left[1 - \exp\left(\dfrac{-1}{\taui}\right)\right]
      \exp\left(\dfrac{1 - \Delta y}{\taui}\right),
 \label{eqn:exponential_model}
\end{eqnarray}
where $\nelec^\mathrm{wp}$ is the number of electrons in the delta function, $\nelec^{\rm bg}$ is the number of electrons in surrounding pixels, $y$ is the distance of the delta function from the readout register, and $\Delta y$ is the number of pixels behind the delta function and away from the readout register, into which the trail is created. The factors in square brackets on the first line come from \autorefb{eqn:cloud_volume}, and the factor in square brackets on the second line ensure conservation of charge, i.e.\ that the total number of trailed electrons equals the number of encountered traps:
\begin{equation}
  \sum_{\Delta y=1}^\infty \nelec^\mathrm{trail}(\Delta y) = \left[
    \left(\dfrac{\nelec^\mathrm{wp} - d}{w - d}\right)^\beta
    - \left(\dfrac{\nelec^{\rm bg} - d}{w - d}\right)^\beta \right]
    y~\sum_i^{\mathrm{species}} \rhoi.
 \label{eqn:exponential_model_integrated}
\end{equation}

\subsubsection{Non-instantaneous charge capture}

If traps gradually capture charge during a timescale $\tau^{\rm cap}_i>0$ (as seen in {\sl Gaia}: \citealt{Short2013}, but not previously in {\sl HST}: \citealt{Anderson2010}), we can no longer separate the processes of release then capture, but must calculate their effects simultaneously.

To calculate the effects of charge capture inside $\fV$, we follow \citet{Lindegren1998}. Defining $\tau^{\rm tot}_i \equiv \tau^{\rm rel}_i\tau^{\rm cap}_i/(\tau^{\rm rel}_i+\tau^{\rm cap}_i)$, the fill fraction after time $t$ for initially empty traps is
\begin{equation}
  f^{\rm e}(t;\tau^{\rm cap}_i,\tau^{\rm rel}_i) = \frac{\tau^{\rm tot}_i}{\tau^{\rm cap}_i} \left(
    1 - \exp\left[-\,\frac{t}{\tau^{\rm tot}_i}\right]\right) \;,
    \label{eqn:slow_capture_traps_empty}
\end{equation}
and for initially filled traps is
\begin{equation}
  f^{\rm f}(t;\tau^{\rm cap}_i,\tau^{\rm rel}_i) = 1 - \frac{\tau^{\rm tot}_i}{\tau^{\rm rel}_i} \left(
    1 - \exp\left[-\,\frac{t}{\tau^{\rm tot}_i}\right]\right) \;.
    \label{eqn:slow_capture_traps_full}
\end{equation}
Therefore, for an initial trap occupancy $0\leqslant f(t_0)\leqslant 1$,
\begin{align}
  f(t_0+t_{\rm dwell}) &= f(t_{\rm 0})\, f^{\rm f}(t_{\rm dwell})
    + (1 - f(t_{\rm 0}))\, f^{\rm e}(t_{\rm dwell}) \\
    &= f(t_{\rm 0}) + 
    \left[ \frac{\tau^{\rm rel}_i}{\tau^{\rm cap}_i+\tau^{\rm rel}_i} - f(t_{\rm 0}) \right]
    \left( 1 - \exp\left[-\,\frac{t_{\rm dwell}}{\tau^{\rm tot}_i}\right]\right) \, .
  \label{eqn:slow_capture_traps_release_and_capture}
\end{align}
In all of these cases, the eventual, steady-state solution would be $f(\infty)=\tau^{\rm rel}_i/(\tau^{\rm cap}_i+\tau^{\rm rel}_i)$. 

There is now ambiguity in the order of operations for `high' traps above $\fV$. Released electrons raise the level of the current watermark during $t_{\rm dwell}$. In our implementation we hold the watermark level fixed, and calculate $f$ only one. This leads to very small differences from the implementation of instantaneous capture, even if $\tau^{\rm cap}_i\rightarrow 0$.

Watermarks are monitored in the same way as \S\ref{sec:traps:instant}.
However, because traps are never completely filled, a much larger number of watermarks remain active.
This increases runtime, but an option to periodically `prune' watermarks that contain very few (e.g.\ less than floating point precision) electrons restores some performance.
This also means the occupancy of these trap species must be stored in an additional watermark array, separate to that for instant-capture traps.

\subsubsection{Inhomogeneous release timescales}

If irradiated CCDs are kept cold, then laboratory measurements \citep{Gow2016,Bilgi2019,Parsons2021} and in-orbit measurements with other missions \citep{Skottfelt2024} suggest that the topological defects in the silicon lattice do not thermally anneal into identical states. Each species can include charge traps with a variety of release times $\tau$, roughly following a log-normal distribution
\begin{equation}
  \rho(\tau) = \rho_i ~ \eta(\tau) = \frac{\rho_i}{\tau \sigma^{\rm rel}_i\sqrt{2 \pi}\, }
    \exp\left[
      -\,\frac{\left( \ln \tau - \ln \tau^{\rm rel} _i\right)^2}{2 (\sigma^{\rm rel}_i)^2}
    \right] \;,
  \label{eqn:continuum_traps_distribution}
\end{equation}
where $\tau ^{\rm rel}_i$ and $\sigma^{\rm rel}_i$ set the mean and width.
A population of such traps can have a variety of occupancy fractions $\phi(\tau)$.

If capture is instantaneous, traps' fill fractions are uniquely specified by the time elapsed $t_{\rm elapsed}$ since they were last full
\begin{equation}
  \phi(\tau; t_{\rm elapsed}) = \exp\left[-\dfrac{t_{\rm elapsed}}{\tau}\right] \;,
\end{equation}
such that the total fill fraction is
\begin{equation}
  f(t_{\rm full}) 
    = \int_0^\infty \eta(\tau) ~ \phi(\tau; t_{\rm elapsed}) ~d\tau \,.
  \label{eqn:continuum_traps_ff_from_te}
\end{equation}
Initially empty traps, which have never been filled, start with $f(\infty)$=$0$. 
After dwell time $t_{\rm dwell}$, traps up to the current watermark get filled because $f(0)=1$, and those above the current watermark end up with fill fraction $f(t_{\rm elapsed}+t_{\rm dwell})$.
We convert back-and-forth between $t_{\rm elapsed}$ and $f$ via \autorefb{eqn:continuum_traps_ff_from_te} or by using a root-finder to solve $f(t_{\rm elapsed}) - f' = 0$ and find $t_{\rm elapsed}(f')$. 
In practice, because elapsed times will always be a multiple of $t_{\rm dwell}$, we tabulate values of $f$ and $t_{\rm elapsed}$ once, at the start of the code, then interpolate for $t_{\rm elapsed}(f)$ during each transfer.

If capture is not instantaneous, then capture does not uniformly reset all traps below the current watermark to full. 
Accurately calculating the functional form of $\phi(\tau)$ would then require the full history of capture and release events: which increases storage requirements and runtime.
Instead, we approximate this as the closest $\phi(\tau)$ that can be derived from a single release event.
To do this, we calculate a new total fill fraction 
\begin{align}
 \!\!\!f(t_{\rm elapsed}+t_{\rm dwell}) 
    &= \int_0^\infty \eta(\tau) ~ \phi(\tau; t_{\rm elapsed}) ~d\tau \\
    &= \int_0^\infty \eta(\tau) \Big[
    \phi(\tau;t_{\rm elapsed})\, f^{\rm f}(t_{\rm dwell};\tau^{\rm cap}_i,\tau) 
    \nonumber \\
    & ~~~~~~~~ + \big(1 - \phi(\tau;t_{\rm elapsed})\big)\, f^{\rm e}(t_{\rm dwell};\tau^{\rm cap}_i,\tau)
    \Big] ~ d\tau .
    \label{eqn:slow_capture_continuum_traps_release_and_capture}
\end{align}
This results in a very good approximation of $\phi(\tau)$ for $\tau>\tau^{\rm cap}$.

The occupancy of trap species with inhomogeneous release must be stored in a separate watermark array to those for other species. Indeed, for these traps, it is also more algorithmically efficient to store $t_{\rm elapsed}$ instead of $f$. For example, after evaluating \autorefb{eqn:slow_capture_continuum_traps_release_and_capture}, we update the value of $t_{\rm elapsed}(\,f(t_{\rm elapsed}+t_{\rm dwell})\,)$.

\subsection{ArCTIc model of ROE} \label{sec:arctic:roe}

\subsubsection{Instantaneous transfer between adjacent pixels} 

\arctic assumes that, at the end of an exposure, a camera's readout electronics (ROE) move electrons instantly from one pixel to the next towards the analogue-to-digital converter, and that electrons are captured by or released from traps only for time $t_\mathrm{dwell}$ while they are stationary between transfers. 
Real CCDs also contain traps in the barriers between pixels, and indeed the three or four individually addressable phases within pixels that keep packets separated from their neighbours.
At a cost of (substantial) increase in runtime, the previous \arctic~v6 could accommodate such complexities, including inter-phase voltages set to shepherd released electrons towards the most appropriate phase \citep{Murray2013}. 
In practice, the trailing produced by traps in multiple, identical phases within a pixel was indistinguishable from the trailing produced by traps in a single phase, but with a different, `effective' density of traps (moving exponentials horizontally is degenerate with rescaling them vertically). 
This is therefore the default behaviour in v7.

Algorithmically, it is more efficient to move traps through a fixed image, rather than to move the image through traps. This is because the arrays required to track a 3D population of traps (watermarking and occupancy; see \S\ref{sec:arctic:traps}) have a much larger memory footprint than an image --- and, as we shall describe, this makes it possible to model the traps in each pixel one-at-a-time.
It is also simpler to model all parallel transfers, then rotate the image by $90^\circ$ and model all serial transfers (using the same algorithm with different parameters, although in {\sl ACS/WFC}, \citealt{Rivers2023} find that serial trap densities are negligibly close to zero), rather than interleaving parallel and serial transfers as in a real CCD.
In general, while the order of operations applied to each pixel is important, the overall order of operations applied to the CCD is not.

\subsubsection{Approximating repeated pixel-to-pixel transfers} \label{sec:express}

Reading out a typical CCD with thousands of rows and columns takes a very large number ($n_{\rm col} n_{\rm row} (n_{\rm col}  + n_{\rm row} + 2) / 2$) of pixel-to-pixel transfers.
The transfers (or, more precisely, the dwell times before each transfer) for parallel readout of a {\it single} column can be represented in an $n_{\rm row} \times n_{\rm row}$ matrix, 
\setcounter{MaxMatrixCols}{20}
\begin{equation}
  E_1 = \begin{bmatrix}
    0 & 0 & 0 & 0 & 0 & 0 & 0 & ~      & 0 & 0 & 0 & 0 & 0 & 0 & 1 \\
    0 & 0 & 0 & 0 & 0 & 0 & 0 & ~      & 0 & 0 & 0 & 0 & 0 & 1 & 1 \\
    0 & 0 & 0 & 0 & 0 & 0 & 0 & ~      & 0 & 0 & 0 & 0 & 1 & 1 & 1 \\
    0 & 0 & 0 & 0 & 0 & 0 & 0 & \dots  & 0 & 0 & 0 & 1 & 1 & 1 & 1 \\
    0 & 0 & 0 & 0 & 0 & 0 & 0 & ~      & 0 & 0 & 1 & 1 & 1 & 1 & 1 \\
    0 & 0 & 0 & 0 & 0 & 0 & 0 & ~      & 0 & 1 & 1 & 1 & 1 & 1 & 1 \\
    0 & 0 & 0 & 0 & 0 & 0 & 0 & ~      & 1 & 1 & 1 & 1 & 1 & 1 & 1 \\
    &  &  & \vdots  &  &  &  &  &  &  &  &  \vdots  &  &  &  \\
    0 & 0 & 0 & 0 & 0 & 0 & 0 & ~      & 1 & 1 & 1 & 1 & 1 & 1 & 1 \\
    0 & 0 & 0 & 0 & 0 & 0 & 1 & ~      & 1 & 1 & 1 & 1 & 1 & 1 & 1 \\
    0 & 0 & 0 & 0 & 0 & 1 & 1 & ~      & 1 & 1 & 1 & 1 & 1 & 1 & 1 \\
    0 & 0 & 0 & 0 & 1 & 1 & 1 & ~      & 1 & 1 & 1 & 1 & 1 & 1 & 1 \\
    0 & 0 & 0 & 1 & 1 & 1 & 1 & \dots  & 1 & 1 & 1 & 1 & 1 & 1 & 1 \\
    0 & 0 & 1 & 1 & 1 & 1 & 1 & ~      & 1 & 1 & 1 & 1 & 1 & 1 & 1 \\
    0 & 1 & 1 & 1 & 1 & 1 & 1 & ~      & 1 & 1 & 1 & 1 & 1 & 1 & 1 \\
    1 & 1 & 1 & 1 & 1 & 1 & 1 & ~      & 1 & 1 & 1 & 1 & 1 & 1 & 1 \\
  \end{bmatrix}
  . \nonumber
\end{equation}
Each row of this matrix represents the location of a pixel in the CCD, with readout proceeding up the page and the top row being the pixel nearest to the readout register (distinction between rows and columns is important because the matrix will not stay symmetric). Electrons initially in that top pixel require only a single parallel transfer, represented by the top-right 1. Entries of 0 are ignored. 
Electrons initially in the $i$th row require $i$ parallel transfers, represented by 1s on the diagonal (the transfer from the original pixel) and all entries to the right (transfers through pixels nearer the readout register).
Electrons in the bottom row must be transferred through all $n_{\rm row}$ pixels. 
The sum of each row is the total number of transfers required for that pixel.

In a real CCD, the transfers represented by matrix elements on the diagonal happen first, and simultaneously; these are followed by transfers in successively lower diagonals.
\arctic reorders these operations, starting instead at the top-left, and proceeding down each column in turn. Thus, the first operation modelled is the (dwell time and) transfer of electrons from the bottom pixel into the one above.
Because each column in this matrix involves the traps in a single pixel, the trap variables can be initialised at the top of the column, and merely updated by capturing or releasing charge as they are moved down the image \citep[for serial transfer, the traps are not emptied between columns, so trailing extends from one image row to the next, c.f.][]{Anderson2024}.
The core algorithm for readout is thus a series of nested loops, over:
parallel then serial transfer, 
image columns, 
`express' passes through each column of this matrix as described below, 
and (optionally) phases inside a pixel. 
Inside the innermost loop, each trap species may capture and/or release charge, updating the pixel and the trap states between watermarks, as detailed in \S\ref{sec:arctic:traps}.

If the number of electrons is much greater than the number of traps, then the volume of the charge cloud changes little between transfers.
Combined with our assumption of the same density of traps in each pixel, the effect of each transfer then becomes essentially the same.
For speed, we can therefore compute the effect of only a single transfer on each pixel and then, {\it before applying it to the image}, multiply that effect by the number of transfers that pixel will undertake. We replace 1s in the matrix by this express number of transfers.
The main exception to this is that the first set of traps encountered by a packet of electrons (those on the diagonal) are usually empty. 
These may capture a significant amount of charge, and initially contain none to release.
By default, we therefore treat this first capture separately, preserving the diagonal entries in a matrix such as
\begin{equation}
  E_3 = \begin{bmatrix}
    0 & 0 & 0 & 0 & 0 & 0 & 0 & ~      & 0 & 0 & 0 & 0 & 0 & 0 & 1 \\
    0 & 0 & 0 & 0 & 0 & 0 & 0 & ~      & 0 & 0 & 0 & 0 & 0 & 1 & 1 \\
    0 & 0 & 0 & 0 & 0 & 0 & 0 & ~      & 0 & 0 & 0 & 0 & 1 & 0 & 2 \\
    0 & 0 & 0 & 0 & 0 & 0 & 0 & \dots  & 0 & 0 & 0 & 1 & 0 & 0 & 3 \\
    0 & 0 & 0 & 0 & 0 & 0 & 0 & ~      & 0 & 0 & 1 & 1 & 0 & 0 & 3 \\
    0 & 0 & 0 & 0 & 0 & 0 & 0 & ~      & 0 & 1 & 0 & 2 & 0 & 0 & 3 \\
    0 & 0 & 0 & 0 & 0 & 0 & 0 & ~      & 1 & 0 & 0 & 3 & 0 & 0 & 3 \\
    &  &  &  \vdots  &  &  &  &  &  &  &  &  \vdots  &  &  &  \\
    0 & 0 & 0 & 0 & 0 & 0 & 1 & ~      & 3 & 0 & 0 & 3 & 0 & 0 & 3 \\
    0 & 0 & 0 & 0 & 0 & 1 & 1 & ~      & 3 & 0 & 0 & 3 & 0 & 0 & 3 \\
    0 & 0 & 0 & 0 & 1 & 0 & 2 & ~      & 3 & 0 & 0 & 3 & 0 & 0 & 3 \\
    0 & 0 & 0 & 1 & 0 & 0 & 3 & \dots  & 3 & 0 & 0 & 3 & 0 & 0 & 3 \\
    0 & 0 & 1 & 1 & 0 & 0 & 3 & ~      & 3 & 0 & 0 & 3 & 0 & 0 & 3 \\
    0 & 1 & 0 & 2 & 0 & 0 & 3 & ~      & 3 & 0 & 0 & 3 & 0 & 0 & 3 \\
    1 & 0 & 0 & 3 & 0 & 0 & 3 & ~      & 3 & 0 & 0 & 3 & 0 & 0 & 3 \\
  \end{bmatrix}
  , \nonumber
\end{equation}
where, in this example, we have replaced up to three transfers with one. Operations with a zero entry are again skipped: run time is nearly a factor 3 quicker. For remaining operations, the effect on the image is multiplied by the number in the matrix (which need not be an integer). Note how the sum of each row in $E_3$ --- the total number of transfers for that pixel --- is identical to the sum of each row in $E_1$, or in any $E_n$.

\begin{figure*}
	\centering
	\includegraphics[
    width=\textwidth, trim={30mm 8mm 18mm 26mm}, clip]{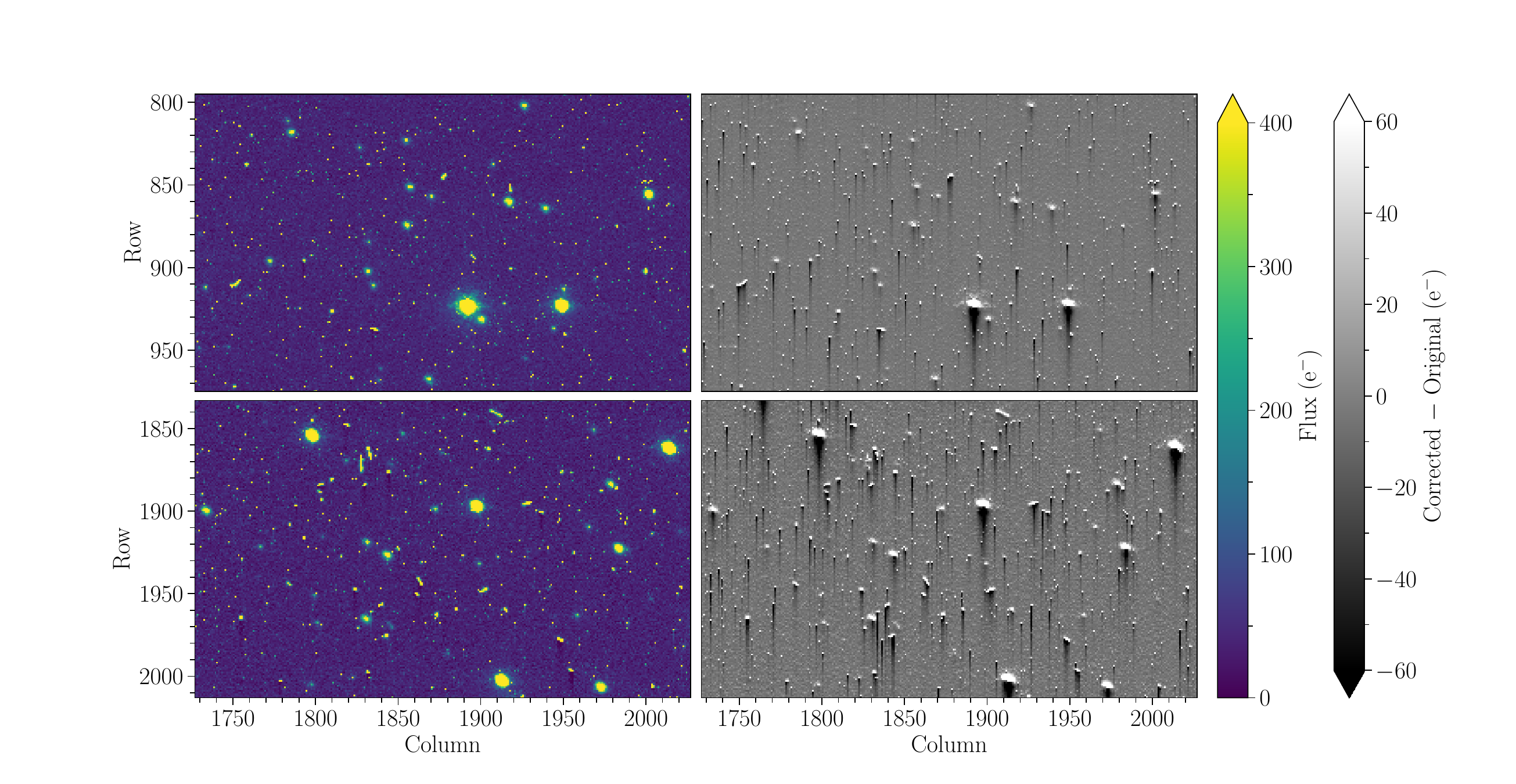}
  \caption{
    The same HST ACS image zoom regions of \autoref{fig:example_image_zooms} after CTI correction with \arctic to remove the trails. 
    The right panels show the difference between the original and corrected fluxes.
	\label{fig:example_image_corrected}}
\end{figure*}

The choice of sparsity in this matrix $E_n$ is a trade-off between accuracy and computation speed.
In practice, we find that the output is consistent to within one electron per pixel if we replace $\sim$$400$ transfers with one. In this case, for a CCD with $2048\times2048$~pixels, $E_{410}$ is the first that has only 5 columns containing off-diagonal elements, and runtime is $\sim$$300$ times faster than $E_1$.
The fastest runtime would use $E_{2048}$, which contains only ones on the diagonal and $1,\ 1,\ 2,\ 3,\ 4\dots2047$ in the rightmost column.

Note this is different from previous algorithms
(and an option in \arctic not to treat first captures separately),
which use a matrix like \!\!
\begin{equation}
  E'_{2048} = 
  \begin{bmatrix}
    1 \\ 2 \\ 3 \\ 4 \\ 5 \\ 6 \\ 7 \\ 8 \\ 9 \\ 10 \\
    \vdots \\ 
    2039 \\ 2040 \\ 2041 \\ 2042 \\ 2043 \\ 2044 \\ 2045 \\ 2046 \\ 2047 \\ 2048
  \end{bmatrix}
  ~~~~~\mathrm{or}~~~~~~
  E'_{512} = 
  \begin{bmatrix}
    1~ & 0~ & 0~ & 0 \\
    2~ & 0~ & 0~ & 0 \\
    3~ & 0~ & 0~ & 0 \\
    \vdots & \vdots & \vdots & \vdots \\
    511~ & 0~  & 0~ & 0 \\
    512~ & 0^{\,\rm r}\!\!  & 0~ & 0 \\
    512^{\,\rm s}\!\! & 1~  & 0~ & 0 \\
    512~  & 2~  & 0~ & 0 \\
    \vdots & \vdots & \vdots & \vdots \\
    512~ & 511~ & 0~ & 0 \\
    512~ & 512~ & 0^{\,\rm r}\!\!  & 0 \\
    512~ & 512^{\,\rm s}\!\! & 1~ & 0 \\
    512~ & 512~ & 2~ & 0 \\
    \vdots & \vdots & \vdots & \vdots \\
    512~ & 512~ & 511~ & 0 \\
    512~ & 512~ & 512~ & 0^{\,\rm r}\!\! \\
    512~ & 512~ & 512^{\,\rm s}\!\! & 1 \\
    512~ & 512~ & 512~ & 2 \\
    \vdots & \vdots & \vdots & \vdots \\
    512~ & 512~ & 512~ & 511 \\
    512~ & 512~ & 512~ & 512
  \end{bmatrix} 
  .
  \nonumber
\end{equation}
The single-column version was independently suggested by \citet{Anderson2010} and \citet{Short2013}; the multicolumn generalisation by \citet{Massey2010}. 
This matrix has even fewer non-zero entries, hence faster runtime. But the problem with the multicolumn version is that (for $E'_{512}$) image rows 513, 1025 and 1537 encounter empty traps (with a matrix entry 1) that capture more electrons than all other pixels. The location of these special pixels depends on the number of columns in the matrix. In this case, an approximation technique leads to residuals with spatially recurring patterns, which is often undesirable.

It is possible to ignore the extra effectiveness of these first captures, and smoothly improve accuracy while retaining the behaviour of $E'_{2048}$ by saving the trap state after entries with a superscript ${\rm s}$, then restoring it after entries with a superscript ${\rm r}$. However, this neglects the additional charge capture into initially empty traps, which is itself a non-linear transformation of the image data.

\subsubsection{Non-standard readout schemes} 

By adjusting the $E$ matrix, it is possible to simulate different readout modes. For example, electrons injected electronically into certain pixels \citep[to mitigate CTI by pre-filling traps;][]{Bushouse2011} all undergo the same number of transfers toward the readout register. The corresponding charge-injection matrix $E_1^{\rm CI}$ contains all 1s, and $E_n^{\rm CI}$ is an $(n_{\rm row}/n)\times n_{\rm row}$ matrix containing all $n$s.

It is also possible to `reverse clock' electrons in the opposite direction to readout. `Pumping' an initially flat image back and forth creates dipoles at the locations of traps, when electrons are captured and released asymmetrically in adjacent pixels \citep{Dawson2008,Skottfelt2024}.

The current code can implement only one of these readout modes at a time, but this could be extended if needed to, for example, track injected charge on top of a standard imaging exposure.

\subsection{Removing CTI trailing} \label{sec:remove_cti}

We have so far described an algorithm to {\it add} CTI trailing to an image. It is not possible to analytically invert the algorithm, and move trailed electrons back to their original pixels. However, we can approximately undo the effects of CTI by considering the trails as a small perturbation on the true image. \autoref{tab:remove_cti} describes how an arbitrarily close approximation of the unknown true image can be obtained, by successively forward-modelling the addition of CTI to the known image-with-trails that is read out from a CCD -- provided of course that the model used to add CTI is sufficiently accurate.

In practice, we use express $E_{n=410}$ and $n_\mathrm{iter}=5$ iterations to correct {\sl ACS/WFC} images, after which further changes are less than one electron per pixel in all tested images. This procedure removes trailed electrons and adds them back to their original bright pixels, as illustrated for a single iteration in \autoref{fig:example_image_corrected}.

\begin{table}
  \begin{center} \begin{tabular}{llr}
    \hline
    Unavailable true image & $I$ & \\
    Available image & $I + \delta$ & (A) \\
    \hline
    Add CTI to A & $I + 2\delta + \delta^2$ & (B) \\
    ${\rm A} + {\rm A} - {\rm B}$ & $I - \delta^2$ & (C) \\
    Add CTI to C & $I + \delta - \delta^2 - \delta^3$ & (D) \\
    ${\rm A} + {\rm C} - {\rm D}$ & $I + \delta^3$ & (E) \\
    \vdots $\;$ (add CTI $n$ times) & \vdots & \\[0.3em]
    ${\rm A} + {\rm X} - {\rm Y}$ & $I \pm \delta^n$ & $\approx I$  \\
    \hline
	\end{tabular} \end{center}
	\caption{
    Iterative correction of CTI trails to recover a true image
    by successively forward-modelling the addition of CTI.
    The known input image is taken as a combination
    of the desired true image, $I$,
    and the unwanted perturbation of the CTI trails, $\delta$.
    \label{tab:remove_cti}}
\end{table}

\subsubsection{Overcorrection of read noise} \label{sec:s+r}

Unfortunately, an {\sl ACS/WFC} image $I$ is a combination of emission from the sky, $S$, which is trailed by CTI, plus readout noise, $R$, a realisation of white, Gaussian noise with rms $\sigma_{\rm rn}$$\sim$\,4\,electrons pixel$^{-1}$ \citep{Desjardins2019} that is added after readout and therefore not trailed. If the combined image $I=S+R$ is corrected for CTI, then trailing could be removed perfectly from $S$ but additional ``trailing'' is incorrectly removed from the read noise component that was not trailed in the first place \citep{Anderson2010}. Just as CTI creates covariance between adjacent pixels, overcorrection of read noise creates spurious anticorrelation.

To mitigate the overcorrection of read noise, we attempt to isolate and correct an estimate of the sky image, $\tilde{S}$, outputting 
\begin{equation}
    I_{\rm output}={\rm correction}(\,I-f_R\tilde{R}\,)+f_R\tilde{R} ,
    \label{eqn:SRfraction}
\end{equation}
where $\tilde{R}$ is an estimate of read noise, and $f_R$ is a fraction, ideally 100\%, that we shall determine later.
Intuitively, it sounds impossible to isolate and subtract one source of noise amongst others. 
But even if our estimate of read noise $\tilde{R}$ is completely wrong, the worst thing we have done is to subtract then add a bit more read noise. If our estimate of read noise $\tilde{R}$ is even partially correct, the overcorrection should be reduced. And since $S$ has been smoothed by CTI trailing, the white read noise component can be modelled with surprising accuracy.

To estimate $\tilde{S}$ we adopt and improve the algorithm of \cite{Anderson2018}\footnote{\url{https://github.com/spacetelescope/hstcal/blob/master/ctegen2/ctegen2.c\#L625}}. This begins with a copy of $I$ then iteratively adjusts the value of each pixel, based on comparisons to the values of its eight surrounding neighbours.
With a target of generating a model $\tilde{R}=I-\tilde{S}$ with pixel variance $\sigma^2_R=\sigma^2_{\rm rn}$ (within $0.01\%$ tolerance, or truncated after a maximum of 200~steps), at iteration $k$, each pixel value $\tilde{S}_k(i,j)$ is changed to
\begin{eqnarray}
\tilde{S}_{k+1}(i,j) = \tilde{S}_k(i,j) + \frac{1}{6} 
\left(
\frac{\Delta_{\rm im}}{\left(1+\frac{2\sigma_{\rm rn}^2}{\Delta^2_{\rm im}}\right)}
+ 
\frac{\Delta_{\rm sm}}{\bigg(1+\frac{18\sigma_{\rm rn}^2}{\Delta^2_{\rm sm}}\bigg)} \,+ \right. \nonumber ~~~~~~~~ \\
\left.
   \frac{\Delta_{\rm {p+}}}{\left(1+\frac{\Delta^2_{\rm {\rm {p+}}}}{2\sigma_{\rm rn}^2} \right)}
 + \frac{\Delta_{\rm {p-}}}{\left(1+\frac{\Delta^2_{\rm {\rm {p-}}}}{2\sigma_{\rm rn}^2} \right)}
 + \frac{\Delta_{\rm {s+}}}{\bigg(1+\frac{\Delta^2_{\rm {\rm {s+}}}}{2\sigma_{\rm rn}^2} \bigg)}
 + \frac{\Delta_{\rm {s-}}}{\bigg(1+\frac{\Delta^2_{\rm {\rm {s-}}}}{2\sigma_{\rm rn}^2} \bigg)}
\right), \!\!\!
 \label{eqn:sriteration2}
\end{eqnarray}
where
\begin{equation}
    \begin{split}
        \Delta_{\rm im} &= I(i,j) - \tilde{S}_k(i,j) \\
        \Delta_{\rm sm} &= \frac{1}{9} \times\!\!\! \sum_{j'=-1}^1\sum_{i'=-1}^1 \! \Big( I\left(i+i',j+j'\right) - \tilde{S}_k\left(i+i',j+j'\right) \Big)\\
        \Delta_{\rm {p+}} &= \tilde{S}_k(i,j+1) - \tilde{S}_k(i,j) \\
        \Delta_{\rm {p-}} &= \tilde{S}_k(i,j-1) - \tilde{S}_k(i,j) \\
        \Delta_{\rm {s+}} &= \tilde{S}_k(i+1,j) - \tilde{S}_k(i,j) \\
        \Delta_{\rm {s-}} &= \tilde{S}_k(i-1,j) - \tilde{S}_k(i,j) .
    \end{split}
\end{equation}
These perturbations are tuned so that the first term keeps $\tilde{S}$ close to $I$, the second term keeps a smoothed version of $\tilde{S}$ close to a smoothed version of $I$, while the other four nudge pixel values closer toward one of their neighbours. We find convergence if these six influences are weighted equally, hence the factor 1/6 in \autorefb{eqn:sriteration2}. The weights suppress large adjustments that would change pixel values by more than $2\sigma_{\rm rn}$.
In the absence of serial CTI (as with ACS), we omit the $\Delta_{\rm mod\_{s}}$ terms and change the prefactor to 1/4, recovering the algorithm of \cite{Anderson2018}. 

Since $\tilde{R}$ is an imperfect estimate of $R$, in practice we achieve better correction by subtracting only a fraction of $\tilde{R}$ in \autorefb{eqn:SRfraction}. While the CALACS implementation hardcodes this as $f_R=0.75$, which works well for some recent data, we find that the optimum value depends on the trap density and the background level. When correcting an image, we measure the sky background level, create and correct a mock image of just sky background (plus shot noise and readout noise), adjusting the fraction $f_R$ of $\tilde{R}$ subtracted in order to optimise some figure of merit (our default option for which is the sum of the pixel-to-pixel covariance matrix by up to $\pm$2~pixels lag). 

This procedure has the useful benefit of also returning a model of the residual covariance  between pixels due to the mismatch between $R$ and $\tilde{R}$. 
It is impossible to represent the covariance everywhere in a science image using a single matrix, because the covariance is not stationary: it depends on the number of electrons in a pixel, and the number of electrons in surrounding pixels that can pre-fill traps.
As the most useful approximation, we compute the covariance matrix in a blank patch of sky.
For faint galaxies, this will be appropriate; for bright stars, it will be dwarfed by the autocorrelation due to shot noise in the sky background, but is nonetheless a valid component.
We calculate a covariance matrix at each corner of each CCD quadrant; since CTI is linear in position, the relevant matrix can then be inferred anywhere in the quadrant using bilinear interpolation.

\section{Measuring CTI in the Hubble Space Telescope} \label{sec:calibration}

In this section, we calibrate the \arctic\ model parameters for, and hence measure the accumulated radiation damage to {\sl HST}'s \acswfc camera. 
We achieve this by exploiting warm pixels: randomly distributed spots in the CCD area, also caused by radiation damage, where charge continuously leaks into the silicon substrate \citep[e.g.][]{Biretta+Kozhurina-Platais2005}.
Charge leakage during an exposure creates `warm' or `hot' pixels (with a brightness depending on the leak current) that appear as isolated delta-functions in the absence of CTI. 
Leakage continues during readout, so the entire column contains a slightly elevated background of spurious charge \citep[the level of which may not correlate exactly with the brightness of the warm pixel itself:][]{Ryon2022}.

We characterise CTI using the `extended pixel edge response' (EPER) trailing of these delta functions.
Were such ideal sources not available, it might also be possible to measure the `first pixel response' (FPR) after negative sink pixels \citep{Guzman2024}, or both FPR and EPER of flat field images into overscan pixels.

\subsection{Finding warm pixels} \label{sec:warm_pixel_finding}

We identify $n_\mathrm{epoch}=178$ `epochs' between late 2002 and early 2025, when {\sl ACS} imaged the extragalactic sky in the F814W band with many exposures of $\sim$500\,seconds each. Such images are routinely acquired for a wide range of science goals, and constitute a substantial fraction of the archive. They are useful for our purposes because they have a similar sky background level ($\nelec^\mathrm{bg}$$\sim$$100$\,electrons) and contain sparse, uncrowded astronomical sources. We permit no undithered duplicate exposures, and no more than two exposures in a set of dithered exposures of a single target. Astronomical sources thus appear at random ($x$, $y$) locations in these data, but the warm pixels persist in their fixed locations (some fraction of warm pixels are removed by annealing, but warm-up events have never been more frequent than monthly). For each epoch, we download 8 to 20 \texttt{\_raw.fits} images taken within a few days of each other, using the Mikulski Archive for Space Telescopes\footnote{\url{https://archive.stsci.edu/}}. 
To convert images to units of electrons, we multiply them by the calibrated gain \citep{Anand2023}, then subtract the monthly superbias image to remove fixed bias structure and dark current accrued during readout \citep{Coe+Grogin2014}. 

We split each image into the four CCD quadrants, reorienting each to have the same parallel and serial clocking directions \citep{Ryon2023}. 
Henceforth, we shall combine data from all four quadrants. 

\begin{figure}
	\centering
	\includegraphics[
    width=\columnwidth, trim={41mm 7mm 4mm 27mm}, clip]{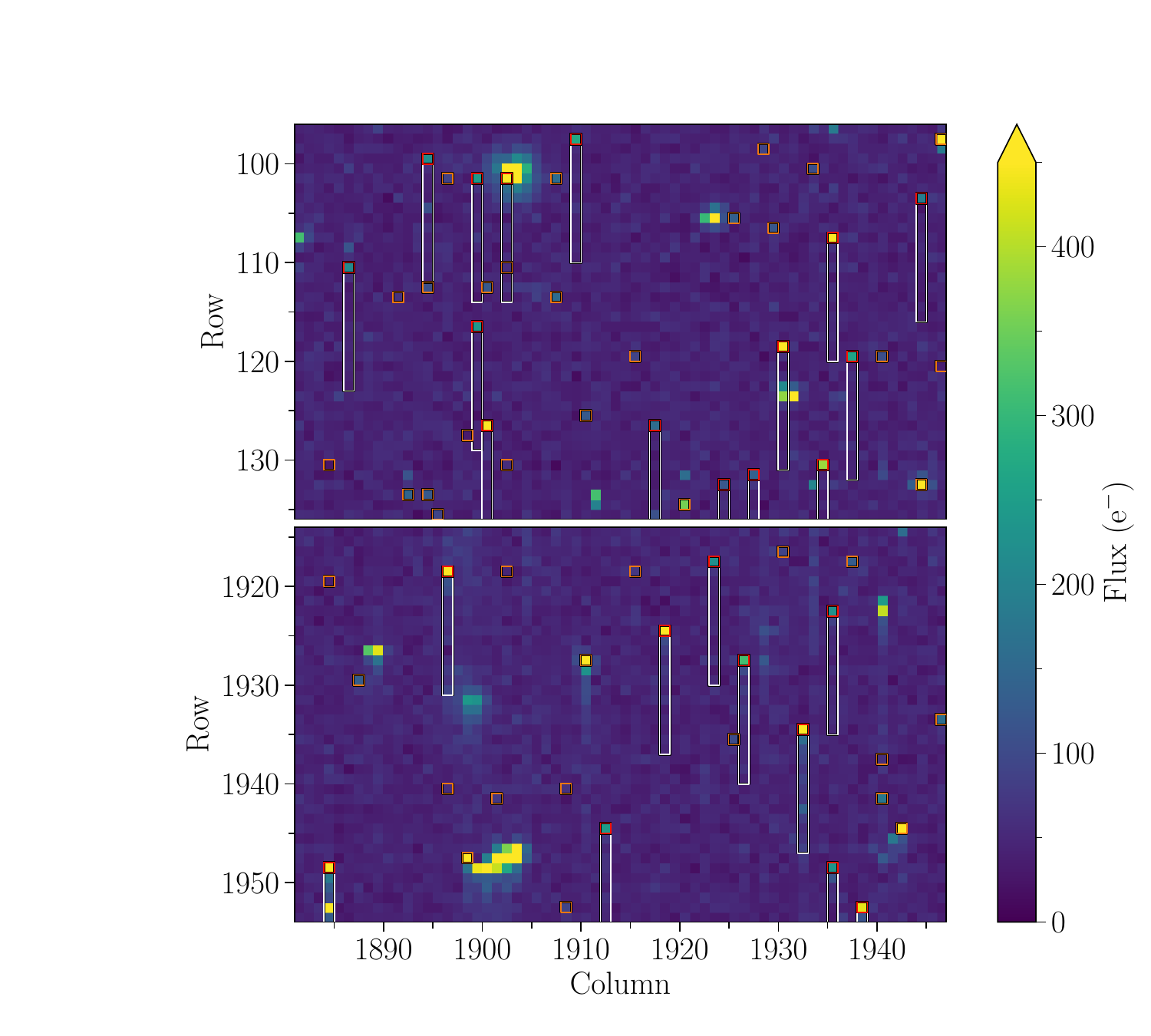}
    \caption{
    An example of identified warm pixels,
    from the same \acswfc image as \autoref{fig:example_image_zooms},
    further zoomed in to the top-middle of the
    two regions shown in \autoref{fig:example_image_zooms},
    near and far from the readout register that is above row~0.
    Orange squares indicate candidate warm pixels, identified in this image alone. Red squares above white rectangles show the warm pixels that were identified consistently
    at the same location in at least 2/3 of the images from this epoch.
    White rectangles confirm the pixels extracted and averaged as CTI trails.
    \label{fig:found_warm_pixels}}
\end{figure}

We identify warm pixels as local maxima of an image that appear consistently at the same pixel in at least two-thirds of the images in an epoch, following the procedure of \citet{Massey+2014}. To exclude astronomical sources, we also require that the value in a warm pixel candidate be greater than the pixels above and below it, and 6 times greater than all four adjacent pixels after smoothing by a $3\times3$\,pixel top hat kernel (\autoref{fig:found_warm_pixels} shows an example selection). To avoid saturated pixels, whose electrons bleed rather than trail into adjacent pixels, we ignore saturated pixels or CCD columns with a median value $3.5\sigma$ higher than other columns. So that we can measure trailing, we also ignore regions within 12 pixels of the top and bottom of the CCD.

\subsection{Measuring trails behind warm pixels} \label{sec:warm_pixel_measure}

We extract the trail behind each warm pixel by taking the values of the 12 pixels below it, minus those of the 12 pixels above it: exploiting symmetry to remove sky background and residual flux from faint astronomical sources below the detection threshold. 
Compared to the sky background, this process appears accurate within typically $\sigma_\mathrm{bg}\approx1$\,electron.
The 12~pixel length of the extracted data is a compromise between capturing the full trail and avoiding contamination from other objects in the image.

\begin{figure*}
  \centering
  \includegraphics[width=\textwidth, trim={8mm 9mm 21mm 8mm}, clip]{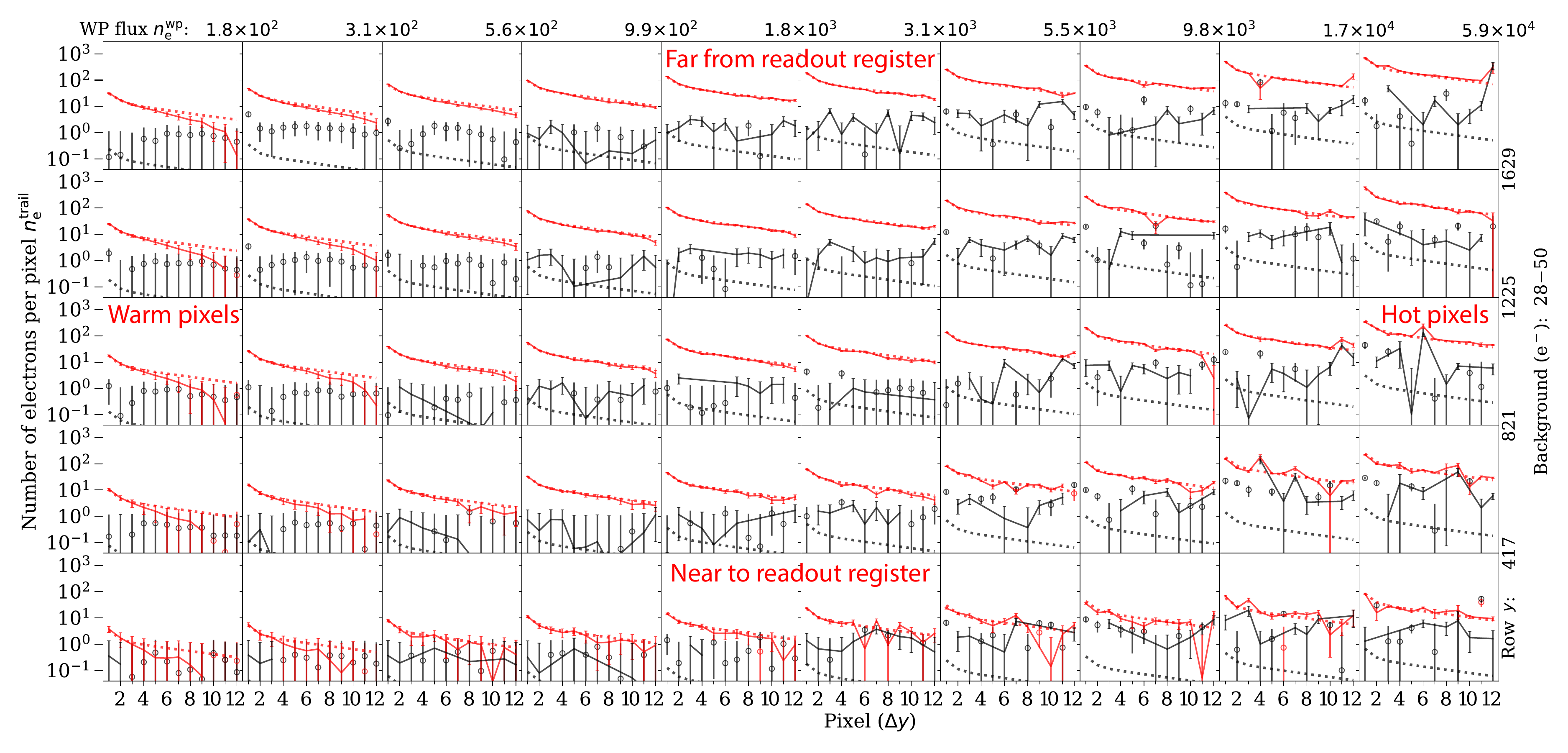}
  \caption{Measured CTI trails $\nelec^\mathrm{trail}(\Delta y)$ in one example epoch on 2024 July 26; we have similar data in each of 178 epochs throughout the lifetime of {\sl ACS}.
  Red points in each panel show the mean value of 12 pixels behind a warm pixel, minus 12 pixels in front; dotted red lines show an \arctic model simultaneously fitted to all panels (to be conservative, we here show a fit with most parameters fixed to values in \autoref{tab:arctic_params} and only $\rhotot$ free; see \S\ref{sec:calibrate_time_evol_physical}). Open circles indicate negative points.
  Each panel averages trails from warm pixels of different brightness $\nelec^\mathrm{wp}$ (increasing left to right, starting at the $\sim$50~electron sky background and separated by values shown at the top) and distance from the readout register $y$ (increasing bottom to top, separated by values shown on the right). 
  Black data points and dotted lines show measurements and the absolute value of the best-fit model (\autoref{eqn:exponential_model_rhotot}) after correction.
  \label{fig:stacked_trails}}
\end{figure*}

We average the trails behind warm pixels in five linearly-spaced bins of distance from the readout register, $y$, and ten logarithmically-spaced bins of  observed warm-pixel flux, $\nelec^\mathrm{wp,obs}$.
The trail amplitude increases with both 
(\autoref{fig:stacked_trails}).
We estimate the uncertainty $\sigma_{\nelec}$ on each data point as the standard deviation of the averaged trails divided by the square root of the number of stacked trails, added to $\sigma_\mathrm{bg}$ in quadrature. Our results are insensitive to the exact value of $\sigma_\mathrm{bg}$, but having a nonzero value eliminates an instability in \citet{Massey+2014}, dealt with there by a fairly arbitrary reweighting scheme. We now understand that uncertainty in background subtraction, leading to occasional negative points in the trail behind warm pixels with the lowest flux, dominated and biased that model fit because the $\sim$100$\times$ more warm pixels in these bins than those with the highest flux led to incorrectly low statistical uncertainty. 

Inferring what the warm pixels looked like before CTI trailing, we also construct 50 model images (one for each panel in \autoref{fig:stacked_trails}). They contain a single bright pixel of amplitude $\nelec^\mathrm{wp}$ surrounded by pixels at the sky background, $\nelec^\mathrm{bg}$. \revone{Following \cite{Bertin1996}, albeit without masking astronomical sources, we estimate the background level} as $2.5\times$ the median minus $1.5\times$ the mean, of the mean flux in all non-discarded columns in a CCD quadrant. The amplitude $\nelec^\mathrm{wp}$ is the mean number of electrons in the observed warm pixels plus the number of electrons in the mean trail: 
\begin{equation}
  \nelec^\mathrm{wp} = \nelec^\mathrm{wp,obs} + \sum_{\Delta y=1}^\infty \nelec^\mathrm{trail}(\Delta y).
 \label{eqn:pre-CTI_amplitude}
\end{equation}
The second term, which adds a few trailed electrons to the model of the pre-CTI warm pixel seems a small effect \citep[and was not done by][]{Massey+2014}, but we shall find that it improves the performance of CTI correction by an order of magnitude. 

A limitation of our analysis is that we take a simple average of the trails within each bin, despite an expectation that the trail amplitude should increase nonlinearly with warm pixel flux, and that the distribution of warm pixel fluxes is nonuniform within a bin. 
However, binning in this way made the data volume manageable and subsequent data analysis feasible in reasonable runtime.

\subsection{Trails are well-fit with instantaneous capture and homogeneous release} \label{sec:trail_fits}

We calibrate the CTI model at each time epoch by repeatedly passing models of the pre-CTI warm pixels through \arctic, and varying the model parameters until the 50 output trails (one for each panel in \autoref{fig:stacked_trails}, shown as dotted red lines) match the 50 observed trails (shown as red data points). 

To optimise \arctic\ model parameters, we use nested sampling code {\tt Dynesty} \citep[version 2.1.0;][]{Speagle2020} to explore the 8-dimensional parameter space $\{\rhoa, \rhob, \rhoc, \taua, \taub, \tauc, \beta, d\}$ of a model with three trap species, with likelihood function $L=\exp{(-\chi^2/2)}$ where $\chi$ is the difference between the trailed model and data, divided by the uncertainty.
We fix $w=84,700$~electrons \citep{Ryon2023} and use uninformative\footnote{Priors are $\rhoa$, $\rhob$, $\rhoc$ flat in [0,5]; $\taua$ Gaussian with $\sigma$=0.2 and $\mu$=0.354(0.541) before (after) the temperature setpoint change, $\taub$ with $\sigma$=2 and $\mu$=4.082(6.466), $\tauc$ with $\sigma$=10 and $\mu$=31.82(57.00); $\beta$ Gaussian with $\sigma$=0.1 and $\mu$=0.556; $d$ flat in [-500,500].}
priors for the other parameters, but force $\taua<\taub<\tauc$ by returning zero likelihood for other regions of parameter space. For fast and accurate convergence, trial and error led us to adopt 500 live points, and random walks \citep{Skilling2006} of 10 steps to find new points with higher log-likelihoods.
We repeat this measurement, separately, at each of 178 epochs during the lifetime of {\sl ACS}.

\begin{figure}
  \centering
  \includegraphics[width=0.89\columnwidth, clip]{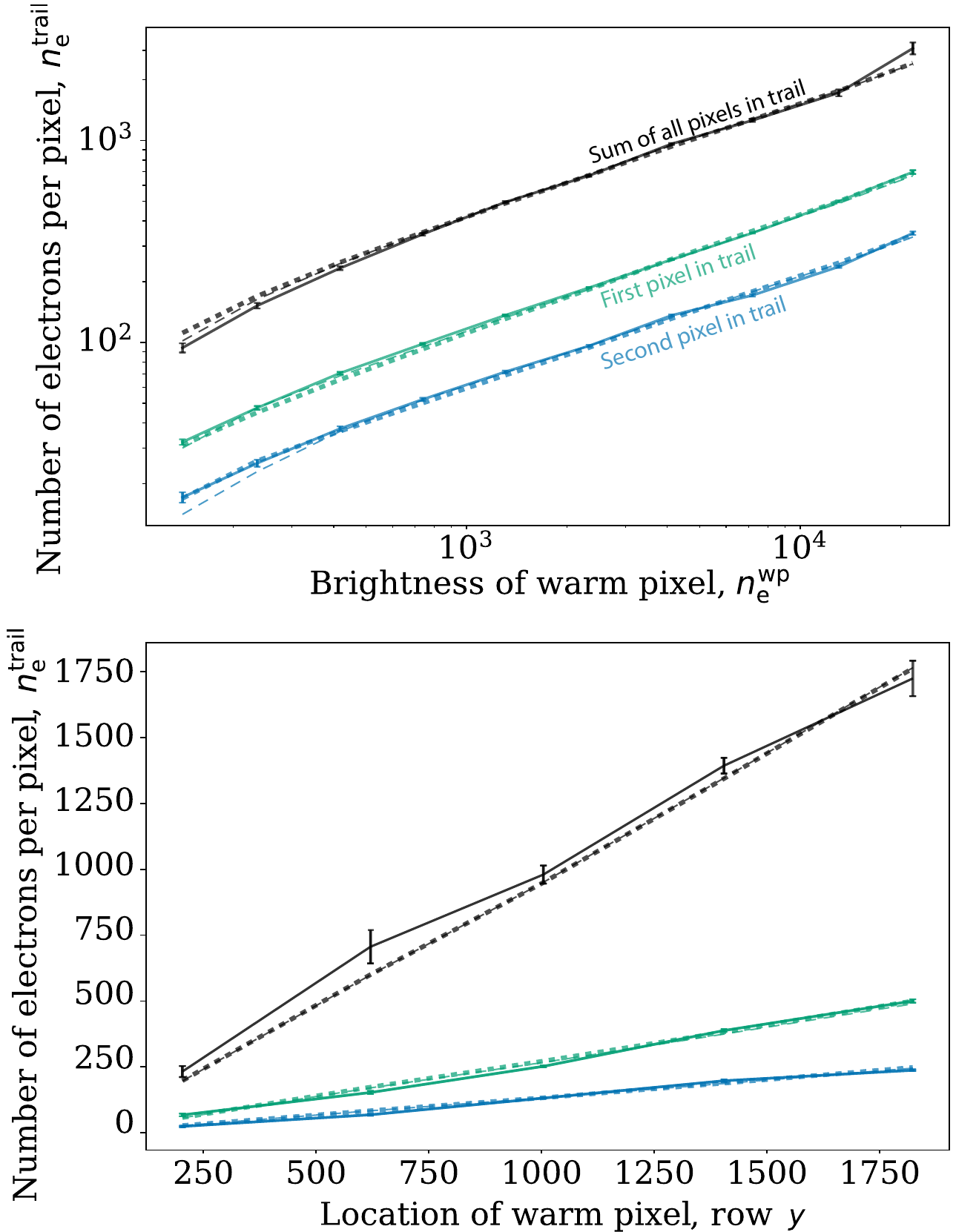}
  \caption{Selected data points from \autoref{fig:stacked_trails}, rearranged to check that the functional form of (the first line of) \autorefb{eqn:exponential_model} is a good fit. Green/blue/black points show the value of the first/second/all pixels in a CTI trail, from trails in the top row of \autoref{fig:stacked_trails} (in panel~a) and in its ninth column (in panel~b). 
  The tenth column is similar but more noisy, and accounts for the upturn at the end of the black points in panel a. \vspace{-5mm}
  \label{fig:trail_checks}}
\end{figure}

In every epoch, the best-fit model of trailing with instantaneous capture and homogeneous release \revone{matches the data in (almost) all bins of warm pixels (\autoref{fig:stacked_trails} dotted red lines}).
This is the first time to our knowledge that trails behind warm pixels at a range of brightnesses have been fitted simultaneously. 
The fit is substantially better than previous analyses that fitted the shape of the trail from only bright warm pixels, then extrapolated that shape to fainter trails (whether they used sums-of-exponentials \citealt{Biretta+Kozhurina-Platais2005,Massey+2010}, empirical \citealt{Anderson2010}, or \arctic\ \citealt{Massey+2014} functions). 
\revone{The only systematic deviation from the model is that trails in the left column of \autoref{fig:stacked_trails} appear slightly steeper than predicted (similarly, the leftmost green and blue points in the top panel of \autoref{fig:trail_checks} are higher than the model, while the leftmost black point is lower). This pattern is consistent across many epochs, recalling \cite{Anderson2024}'s observation in dark exposures that charge clouds with few electrons are preferentially exposed to traps with fast release times.
This could be modelled in \arctic via a different value of $\beta$ for each trap species.
In our data, however, it is difficult to measure the gradient of trails only just above a noise floor --- and the difference in slope disappears if the (pre-subtracted) background level is lowered within $\sigma_\mathrm{bg}$ uncertainty. 
Since our estimated background may include a spurious contribution from unmasked astronomical sources (see \S\ref{sec:warm_pixel_measure}), our data neither support nor refute a species-dependent distribution of traps.}

We find no evidence in {\sl HST}/{\sl ACS} data for the complex trapping behaviour discussed in \S\ref{sec:arctic:traps}. To assess the quality of different model fits, we calculate the Bayesian Information Criterion (BIC). Any figure of merit like this is noisy for individual epochs, so we fit its behaviour over the lifetime of {\sl ACS}. We do not simply average BIC between epochs because no model is perfect, so BIC inevitably grows (the fit gets worse) over time, as the trail amplitude increases. For the simplest model with instant capture and uniform release, we find $\mathrm{BIC}=2985+0.1399\,(t-56,\!000)$ where $t$ is the Modified Julian Date. Allowing delayed charge capture achieves $\mathrm{BIC}=3195$ at MJD 56,000, with best-fit values of $\tau^\mathrm{cap}=\{0.02^{+0.41}_{-0.02},0.001^{+0.006}_{-0.001},0.14^{+0.82}_{-0.14}\}$ all consistent with zero, where the prior is bounded. Allowing continuum charge release achieves $\mathrm{BIC}=2991$ at MJD 56,000, with best-fit values of $\sigma^\mathrm{rel}=\{0.2^{+0.9}_{-0.2},0.006^{+0.223}_{-0.006},30^{+50}_{-30}\}$ again all consistent with zero. Since the simplest model with instant capture and \revone{homogeneous release} is thus both quantitatively best and computationally fastest, we henceforth adopt it for the rest of this paper.

For comparison with earlier work, a sum-of-exponentials model of trail profiles (\autoref{eqn:exponential_model}) achieves $\mathrm{BIC}=3076$ at MJD 56,000 but, when passed through \arctic, its best-fit parameters produce trails biased consistently 5--7\% low. Omitting the second term in \autorefb{eqn:pre-CTI_amplitude}, as for example done by \citet{Massey+2014}, yields $\mathrm{BIC}=3100$ at MJD 56,000: an increase that suggests that the sum-of-exponentials model is not as good as our new, more sophisticated implementation.

\subsection{Time evolution of CTI model parameters} \label{sec:calibrate_time_evol}

\begin{figure}
  \centering
  \includegraphics[width=\columnwidth, trim={36mm 0mm 0mm 0mm}, clip]{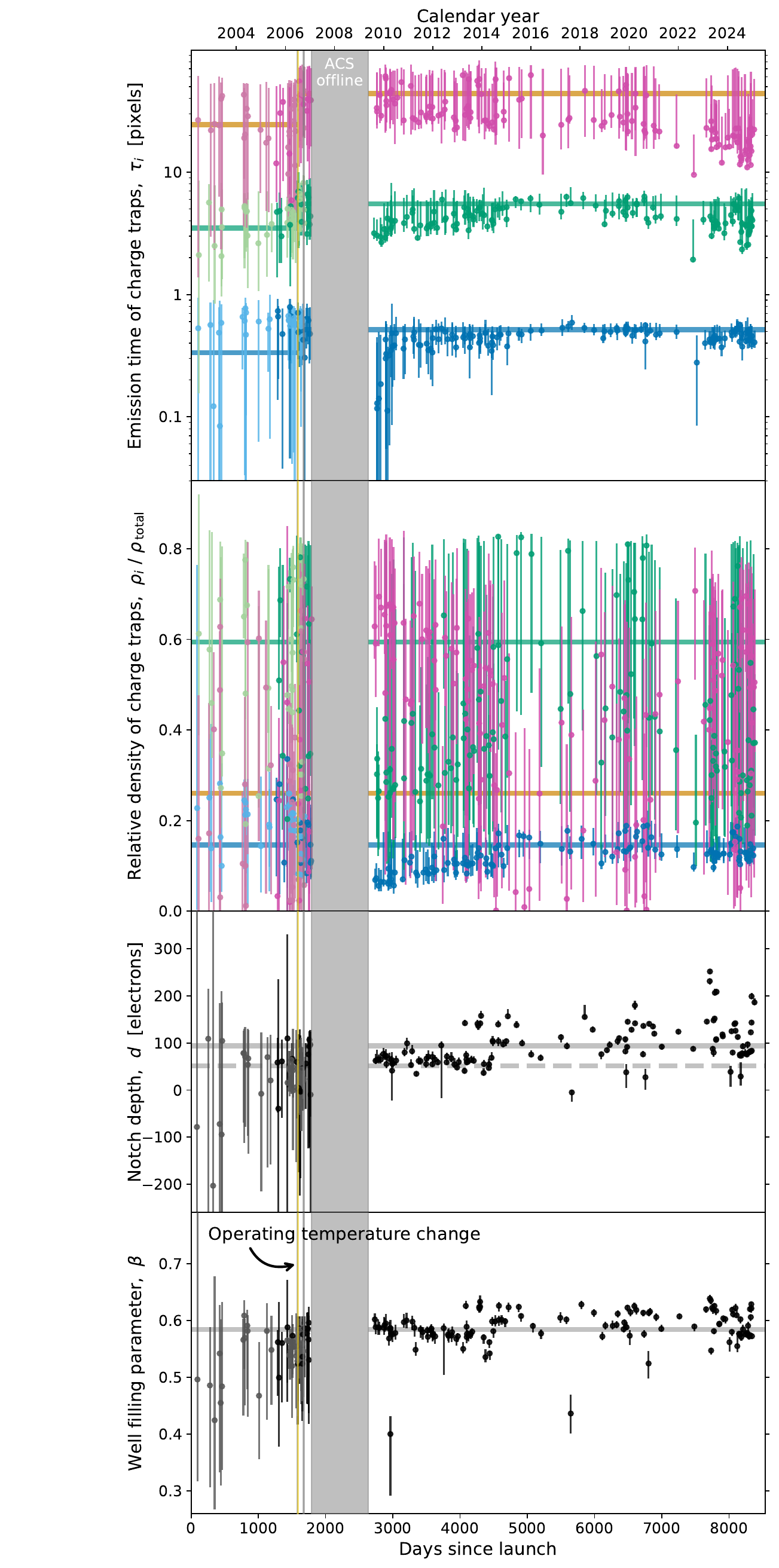}\vspace{-3mm}
  \caption{
  Best-fit values of \arctic parameters concerning device physics, simultaneously fitted from individual epochs of {\sl HST ACS/WFC} data. Blue, green and purple points correspond to trap species $a$, $b$ and $c$. Solid-state theory suggests that all parameters ought to be constant except traps' characteristic emission times $\tau$, which depend on the CCD operating temperature. Light- (dark-)coloured points show measurements from imaging data acquired with commanded gain=1 (2), which has no measurable effect on results as expected. Vertical bands indicate times when {\sl ACS} was offline, with the gold band indicating when the temperature set-point of the CCD was lowered.
  \label{fig:raw_time_evol_all}
  \vspace{-10mm}}
\end{figure}

To correct science data acquired by {\sl HST ACS/WFC} at any time, we need to interpolate between the best-fit values of \arctic\ model parameters obtained in each of the 178 epochs (\autoref{fig:raw_time_evol_all}, and top panel of \autoref{fig:raw_time_evol_rho} --- which reassuringly resembles Figure~15 of \citealt{Ryon2024}). 
Our goal is to find a physically plausible function of time for each parameter that minimises residuals. 
Ideally, we would fit time-varying model parameters simultaneously to measurements in every epoch, using Bayesian Hierarchical Modelling or Expectation Propagation to find the maximum likelihood and {\it also} explore the full posterior. Unfortunately, we have not managed to get chains to converge across this very broad parameter space.
However, there is considerable freedom to reduce the dimensionality of the fit by fixing the values of some parameters, because measurements within one epoch are highly degenerate: for example, $\rhoi/\rhotot$ and $\tau_i$ are anticorrelated in the way expected from fitting decaying exponentials to a finite window of data. 
We therefore consider physical intuition.

\subsubsection{Model parameters concerning device physics}
\label{sec:calibrate_time_evol_physical}

Solid-state theory suggests that characteristic trap emission times should be constant, $\taui\propto T^{-2}e^{\Delta E_i/kT}$ \citep{ShockleyRead1952,Hall1951} if temperature $T$ is controlled\revone{, where $\Delta E$ is the band gap energy}.
We thus adopt the linear average of these parameters across all epochs, factoring in knowledge that the {\sl ACS} detector temperature setpoint was lowered from $-77^\circ$C to $-81^\circ$C in July 2006 \citep[MJD 53920;][]{Sirianni+2006,Mack+2007}, and that the band gap energies $\Delta E=\{0.31\,\mathrm{eV}, 0.34\,\mathrm{eV}, 0.44\,\mathrm{eV}\}$ of E-centre complexes that are likely responsible for the traps would thus lengthen the release time of the fast, medium, and slow species by factors 1.542, 1.584, and 1.796 respectively \citep{Hopkinson2001,Massey2010}. 

The relative density of each trap species, $\rhoi/\rhotot$, should depend on the (constant) amounts of impurities in the CCD silicon, and the energy spectrum of radiation (which might vary over a Solar cycle).
In principle, traps of different species could be created at different rates during the Solar cycle. 
Empirically, there is marginal statistical evidence for this: e.g.\ $\mathrm{d}(\rhoa/\rhotot) /\mathrm{d}t =(2.8\pm5.9)\times10^{-6}$, but this is not compelling given the high levels of noise \revone{in measurements from individual epochs. The fitted parameters are also highly degenerate with each other (e.g.\ $\rhoi$ with $\taui$, or the different $\rhoi$s with each other).}
We therefore interpret any time variation as statistical noise or covariance, and fix the relative densities of trap species to their mean values over all epochs.

Solid-state theory also suggests parameters $d$ and $\beta$, which describe the size of a cloud of electrons, should also be constant. At late times $t>4000$~days since launch when the signal is cleanest, we measure mean $\beta=0.584\pm0.013$ and $d=94.1\pm2.5$. 
A possible step change in these parameters near $t=4000$~days could be real; it could be influenced by the end of prominent extragalactic surveys that provided extremely uniform data at earlier times; or it could be noise. Both parameters are degenerate with $\rhotot$, and $d$ is the least well constrained of all the parameters, because it controls behaviour below the sky background level in most epochs.

\begin{table}
\centering
\caption{Best-fitting \arctic model parameters, assuming all parameters except $\rho_\mathrm{total}$ are constant over time (also shown as horizontal lines in \autoref{fig:raw_time_evol_all}). 
To convert to physical units, multiply $\tau_i$ by {\sl ACS} clocking rate 3212\,$\muup$s/pixel, and divide $\rho_i$ by pixel volume $\sim$15$\times$15$\times$12.6\,$\muup$m$^3$/pixel.}
\label{tab:arctic_params}
  \begin{tabular}{|l|l|l|} 
    \hline
    Parameter & Units & Value \\
    \hline
    $\beta$ & ~ & 0.584 \\
    $d$ & electrons & 51.1 \\ 
    $\tau_a$, $\tau_b$, $\tau_c$ & pixel & 0.334, 3.491, 24.455 (before June 2006) \\
    ~ & ~ & 0.515, 5.531, 43.924 (after June 2006) \\
    $\rho_a$, $\rho_b$, $\rho_c$ & pixel$^{-1}$ & 0.146, 0.594, 0.260 $\times~\rho_\mathrm{total}$ \\
    \hline
  \end{tabular}
\end{table}

\begin{figure}
  \centering
  \includegraphics[width=\columnwidth, trim={0mm 0mm 0mm 0mm}, clip]{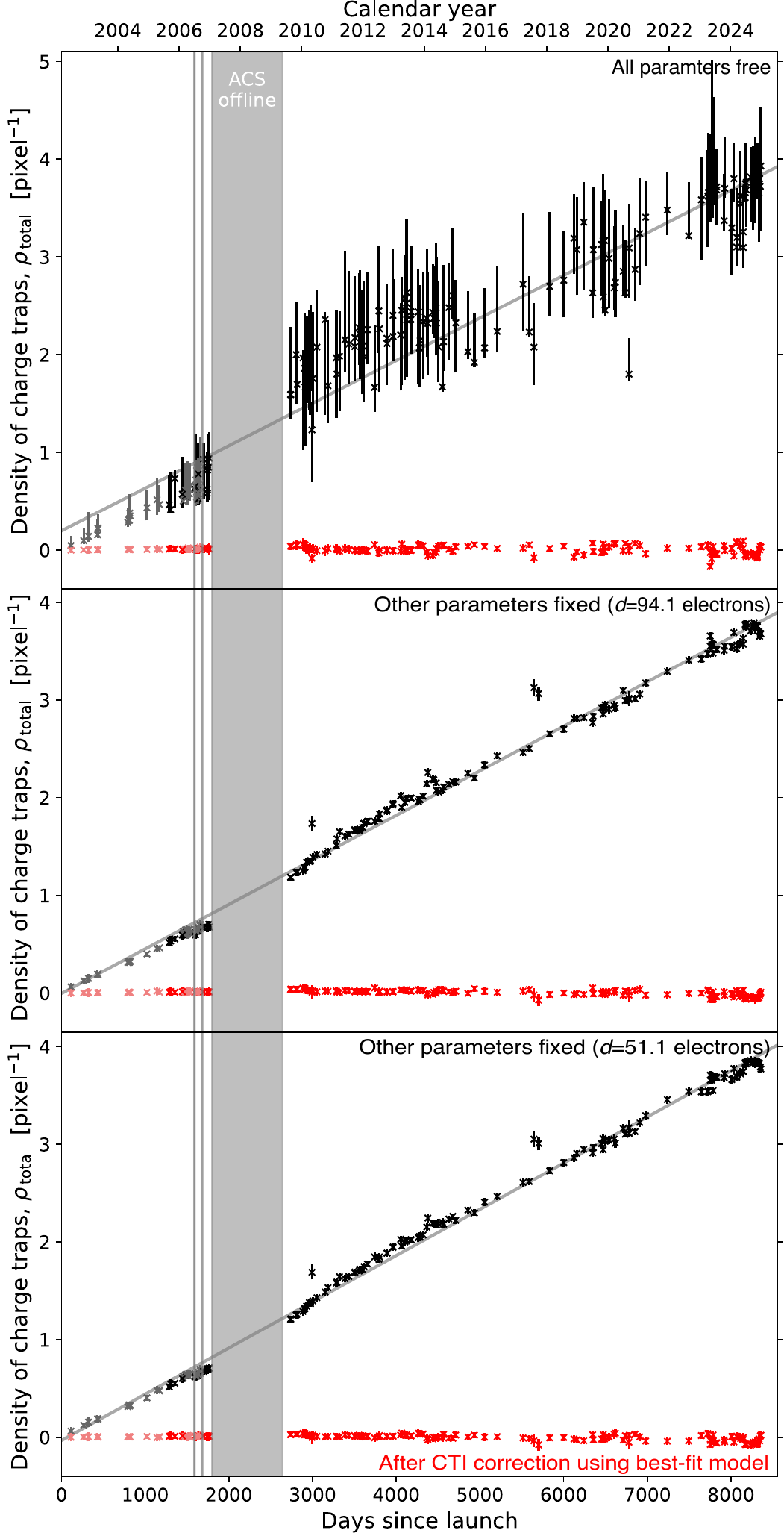}
  \caption{
  The final \arctic model parameter, for the total density of traps, $\rhotot=\rhoa+\rhob+\rhoc$. Black points show values measured from individual epochs of {\sl HST ACS/WFC} data, before CTI correction. Red points show the effective density of traps, measured after CTI correction using model parameters corresponding to the black points from that epoch. The top panel displays large scatter because all 8 model parameters are fitted simultaneously, and they are degenerate. Other panels display less scatter because all parameters other than $\rhotot$ are fixed, to the mean best-fit values across all epochs.
  \label{fig:raw_time_evol_rho}}
\end{figure}

We hold fixed all these physically-motivated parameter values, while \textit{re}fitting at each epoch the final free parameter, $\rhotot$. 
As confirmation that the device model represents the CCDs well, this actually enhances the appearance of \autoref{fig:stacked_trails} at most epochs, by improving the fit the first couple of pixels in trails behind faint warm pixels at the expense of the later ones (dashed lines in \autoref{fig:trail_checks} show the fit with all parameters free; dotted lines show the fit with only $\rhotot$ free). Combining data from all epochs has found a solution with similar or better likelihood, by beating down measurement noise on physically meaningful parameters.
Moreover, breaking the degeneracy between parameters dramatically reduces scatter in $\rhotot(t)$ about linear evolution (middle panel of \autoref{fig:raw_time_evol_rho}).
Closer inspection still reveals correlation between $\rhotot$ and the sky background level at different epochs: inaccuracy in the well-filling model near the sky background is adding either noise or bias to measurements of trap density. Allowing $d$ to vary while we refit $\rhotot$ at each epoch, if $d=51.1$~electrons, we find 50.2~per cent less (and locally minimum) scatter of $\rhotot(t)$ from linear evolution, with trends that are visually more distinct (bottom panel of \autoref{fig:raw_time_evol_rho}). 

We infer this value of $d$ to be better representative, and list it with other final parameters in \autoref{tab:arctic_params}.
Since our data always has more than $51.1$~electrons in every pixel, parameter $d$ is really controlling an interpretation of the macroscopic size of a charge cloud containing a small number of electrons. It should not necessarily be interpreted as evidence for a supplementary buried channel in the e2v CCDs, which was intended during manufacture but remains controversial (L.\ Smith priv.\ comm.).

Note that scatter in $\rhotot(t)$ could alternatively be reduced by 18~per~cent by keeping $d=94.1$~electrons and setting $\beta=0.535$. However, this affects trails behind all warm pixels rather than just those near the sky background level, and substantially degrades the fits to trails behind hot pixels like those towards the right hand side of \autoref{fig:stacked_trails}. We therefore discard this option.

\subsubsection{Model parameters concerning the radiation environment}
\label{sec:calibrate_time_evol_radiation}

\begin{figure*}
	\centering
	\includegraphics[
    width=0.96\textwidth, trim={0mm 0mm 0mm 0mm}, clip]{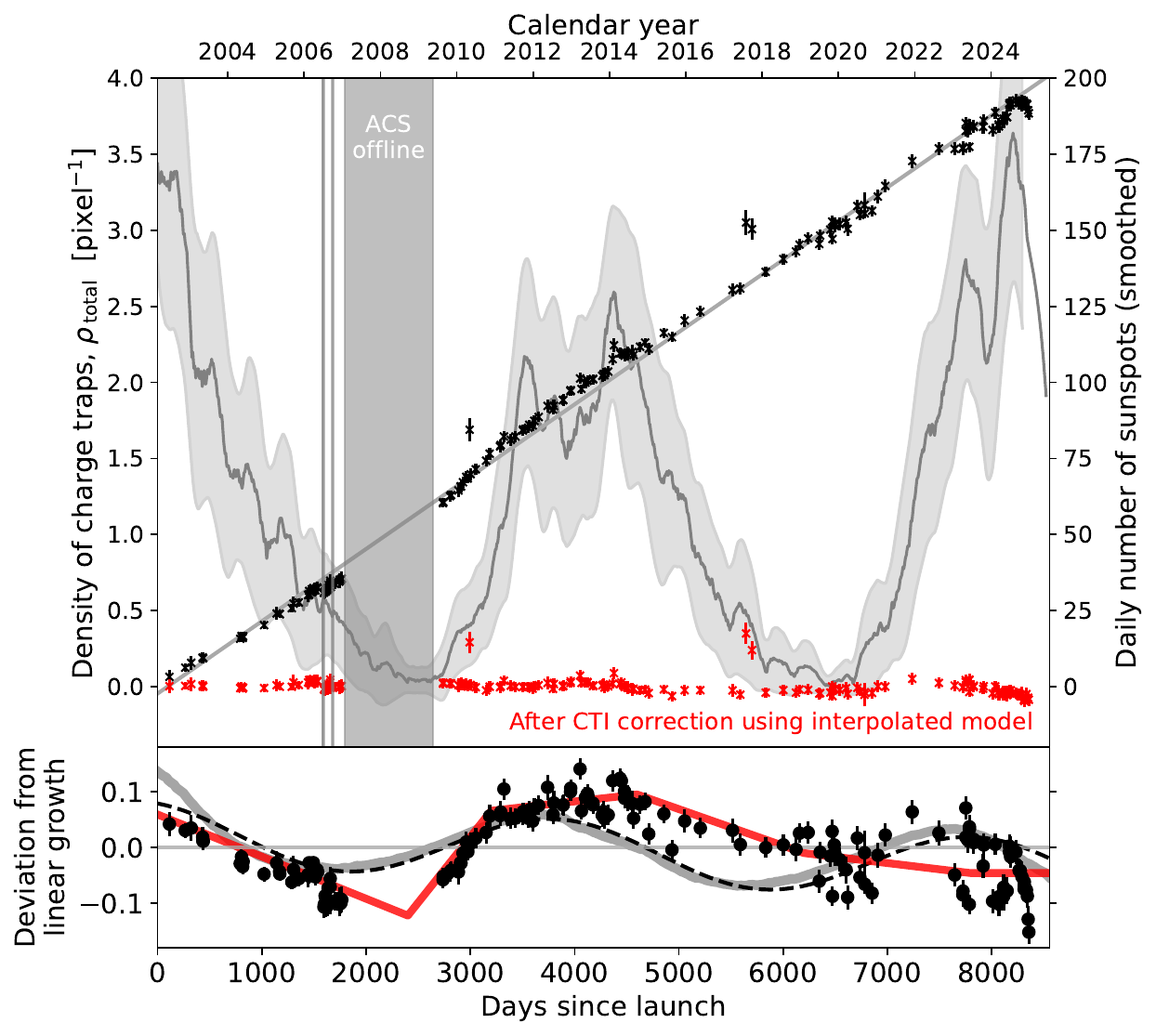}
  \caption{
    The total density of charge traps in {\sl HST} \acswfc as a function of time since launch on 2002 March 1.
    Black points show the measurements from individual epochs.
    Deviations from a linear growth of traps clearly correlate with the Solar cycle, indicated by the light grey shaded region.
    Proposed models of the growth of traps are highlighted in the bottom panel: linear growth (thin grey line), sunspots (thick grey line), sinusoidal (black dashed line), and piecewise linear (red line).    
    Red points show the effective density of charge traps remaining after CTI correction with the piecewise linear model.
	\label{fig:density_evol}}
\end{figure*}

Interpolating measurements of the total number of accumulated traps between epochs, while all other CTI model parameters are held fixed, has best-fit linear evolution \begin{equation}
    \rhotot=(4.7550\pm0.0038)\times10^{-4}~t-(0.0462\pm0.0018),
    \label{eqn:rhotot_linear}
\end{equation}
where $t$ is in units of days since launch. 
The number of intrinsic traps or manufacturing defects, $\rhotot(t=0)$, was much smaller than the number that have accumulated in orbit, as hoped for with an expensive device.

Spatially subdividing CCD quadrants is finally possible when combining the signal-to-noise of many calibration epochs. For a linear fit equivalent to that in \autorefb{eqn:rhotot_linear}, the third of warm pixels nearest the serial register yield a growth rate of traps per day $\rhototdot=(4.7540\pm 0.0039)\times10^{-4}$; the middle third $(4.7531\pm 0.0039)\times10^{-4}$; and the farthest third $(4.7538\pm 0.0039)\times10^{-4}$. The absence of a monotonic spatial gradient of inferred parallel trap density indicates that the small but growing serial CTI in \acswfc \citep{Ryon2024} is not mixing with parallel trails at a sufficient level to cause mis-calibration of parallel CTI (unlike in other cameras, see Nightingale et al.\ in prep.).

Deviations from linear evolution of $\rhotot(t)$ are clearly correlated with Solar activity (\autoref{fig:density_evol}). We initially found it curious that the excess number of sunspots at a given time roughly matches the accumulated damage $\rhotot(t)$ rather than, say, the rate at which damage is being accumulated, $\rhototdot(t)$. This is because radiation in LEO is a combination of both Solar energetic particles and galactic cosmic rays, which are reduced by the Solar wind \citep{Kilifarska2020}. Moreover, sunspots do not cause Solar radiation: rather both are symptoms of the Sun's migrating magnetic structure. The open solar flux (the amount of magnetic flux extending from the Sun into interplanetary space) lags behind sunspot activity by $\sim$324~days, and the neutron flux at the top of the Earth's atmosphere lags by $\sim$100 to 600~days in different Solar cycles \citep{Koldobskiy2022,Tahtinen2024}. Particles can then be trapped for further weeks or months in the radiation belts around Low Earth Orbit \citep{Hands2018,Matthia2023}.

A model involving the daily number of sunspots, $n_{\rm sunspot}(t)$ as tabulated by \cite{sidc}, provides a best-fit evolution (thick grey curve in the bottom panel of \autoref{fig:density_evol})
\begin{eqnarray}
    \rhotot(0)=0.0898\pm0.0045~;   ~~~~~~~~~~~~~~~~~~~~~~~~~~~~~~~~~~~~~~~~~~~~~~~~~~\, \\
    \rhototdot(t)=(5.5738\pm0.0098)\times10^{-4} ~~~~~~~~~~~~~~~~~~~~~~~~~~~~~~~~~~~~~~ \nonumber \\
    -(3.6802\pm0.0334)\times10^{-6}~[n_{\rm sunspot}(t-t_{\rm lag})]^{0.8309\pm0.0020},
    \label{eqn:rhotot_sunspot}
\end{eqnarray}
where $t_{\rm lag}=430^{+4}_{-5}$~days. 
Within uncertainties, this function is consistent with, and has almost identical goodness of fit as a sinusoidal model (dashed black curve in \autoref{fig:density_evol})
\begin{eqnarray}
    \rhotot(t)=(4.6742\pm0.0046)\times10^{-4}~t-(0.0191\pm0.0020) ~~~\nonumber \\
    +\,(0.05514\pm0.001533 )\cos{\frac{2\pi(t+211.2\pm17.3)}{11\times365.25}}~.
    \label{eqn:rhotot_sinusoidal}
\end{eqnarray}
A sinusoidal model with different parameters also roughly fits damage to Gaia at the Earth-Sun Lagrange point L2 (Claudio Pagani, priv.\ comm.). The phase and relative amplitude of our model also roughly match the growth rate of sink pixels \citep[Figure 3 in][]{Guzman2024}.

Assuming that $\rhototdot$ is a good proxy for the intensity of the local radiation environment, we differentiate \autorefb{eqn:rhotot_sinusoidal} and propagate uncertainty to obtain
\begin{eqnarray}
    \rhototdot(t)=(4.6742\pm0.0046)\times10^{-4} ~~~~~~~~~~~~~~~~~~~~~~~~~~~~~~~~~~~~ \nonumber \\
    \times\left[1-(0.1844\pm0.0051)\sin{\frac{2\pi(t+211.2\pm17.3)}{11\times365.25}}\right].
\end{eqnarray}
For the measurements of $\rhotot$ with $d=91.1$, the phase shift increases to $218.3\pm8.0$~days and the modulation to $0.2251\pm0.0047$. 
This increase is far less than the increase in overall scatter of $\rhotot$ with $d=91.1$ but, given the sytematic uncertainty about possible bimodality of $d$ in \autoref{fig:raw_time_evol_all}, we increase error bars to accommodate this, and conclude $18.5^{+4.5}_{-0.5}$~per cent modulation in the intensity of the LEO radiation environment during a Solar cycle, peaking $t_{\rm lag}=430^{+11}_{-5}$~days after Solar minimum.

\begin{table*}
  \begin{center} 
  \begin{tabular}{|l|l|l|r@{${}\pm{}$}l|r@{${}\pm{}$}l|r@{${}\pm{}$}l|r@{${}\pm{}$}l|} 
    \hline 
    \multicolumn{3}{l}{How \arctic\ model parameters are fitted / interpolated} & 
    \multicolumn{2}{c}{$\left\langle\rhotot^\mathrm{corr}\right\rangle$ [pixel$^{-1}$]} &
    \multicolumn{2}{c}{$\left\langle\rhotot^\mathrm{corr}/\rhotot^\mathrm{\meas}\right\rangle$} & 
    \multicolumn{2}{c}{$\left\langle\mu_{\nelec^\mathrm{trail}}^\mathrm{corr}\right\rangle$ [electrons]} &
    \multicolumn{2}{c}{$\left\langle\mu_{\nelec^\mathrm{trail}}^\mathrm{corr}/\mu_{\nelec^\mathrm{trail}}^\mathrm{\meas}\right\rangle$} \\
    \hline 
    \multicolumn{2}{l}{All parameters optimised at every epoch} & Fig.~\ref{fig:raw_time_evol_rho}a &
 0.00549 & 0.00283 &
 0.00650 & 0.00123 &
 0.19216 & 0.01925 &
 0.02443 & 0.00258 \\
    \multicolumn{2}{l}{Fixed ($d=94.1$) except $\rhotot$ optimised at each epoch}   & Fig.~\ref{fig:raw_time_evol_rho}b & 
 0.00346 & 0.00188 &
 0.00708 & 0.00092 &
 0.28307 & 0.02450 &
 0.03575 & 0.00435 \\
    \multicolumn{2}{l}{Fixed ($d=51.1$) except $\rhotot$ optimised at each epoch}   & Fig.~\ref{fig:raw_time_evol_rho}c & 
-0.00150 & 0.00213 &
 0.00492 & 0.00096 &
 0.22940 & 0.01380 &
 0.02886 & 0.00351 \\
    Fixed and $\rhotot(t)$ linear growth & $\chi^2_r=14.00$ & Fig.~\ref{fig:density_evol} & 
-0.00104 & 0.00624 &
 0.00846 & 0.01222 &
 0.23446 & 0.03217 &
 0.02274 & 0.00619 \\
    Fixed and $\rhotot(t)$ from sunspots & $\chi^2_r=10.18$ & Fig.~\ref{fig:density_evol} & 
 0.00259 & 0.00581 &
-0.00440 & 0.00453 &
 0.25034 & 0.02960 &
 0.01113 & 0.01237 \\
    Fixed and $\rhotot(t)$ sinusoidal & $\chi^2_r=10.23$ & Fig.~\ref{fig:density_evol} & 
-0.00136 & 0.00579 &
-0.00820 & 0.00356 &
 0.23047 & 0.02996 &
 0.01248 & 0.00658 \\
    Fixed and $\rhotot(t)$ piecewise linear & $\chi^2_r=~~4.40$ & Fig.~\ref{fig:density_evol} & 
-0.00137 & 0.00380 &
 0.00507 & 0.00225 &
 0.23360 & 0.02240 &   
 0.02842 & 0.00412 \\
    \hline 
    \end{tabular} 
  \end{center}
  \caption{Fractional and absolute residuals after CTI correction of the mean effective density of charge traps and the mean value of trailed pixels, both of which should be zero if the correction were perfect. Uncertainties are standard errors on the mean. To assess performance, these should be compared to mean values before correction of $\langle\rhotot^\mathrm{\meas}\rangle=2.070\pm0.091$ and $\langle \mu_{n^\mathrm{trail}}^\mathrm{\meas} \rangle=8.391\pm0.350$, or approximately twice those values by 2025. Rows 4 to 7, with $\chi^2$ values indicating the goodness-of-fit of $\rhotot(t)$, show different attempts to interpolate \arctic model parameters between the finite number of epochs when they were measured.} 
  \label{tab:results}
\end{table*}

We can find no model involving only Solar activity that captures {\it all} the behaviour of $\rhotot(t)$ --- in particular a decrease in the rate of damage in 2024/5, as if less radiation is reaching LEO during the current Solar cycle than the previous cycle. Because any unmodelled fluctuations in parameter $\rhotot$ will translate into residual, uncorrected CTI, we finally adopt a more flexible but purely empirical approach to time-series interpolation. The best-fit continuous, piecewise linear model (thick red curve in \autoref{fig:density_evol}) is
\begin{equation}
    \rhotot(t)=\rho_\mathrm{launch} + \sum_{i=1}^{6}
    \left[ \min{\big(\max{(t,t_i)},t_{i+1}\big)} -t_i \right] 
    ~\dot{\rho}_i~,
    \label{eqn:rhotot_pwlinear}
\end{equation}
where $\rho_\mathrm{launch}=0.0132\pm0.0004$~traps pixel$^{-1}$, and the gradient changes at times $t_i=\{0, 2400, 3200, 4600, 6200, 7800\}$~days, to $\dot\rho_i=\{3.9995 \pm0.0020, 7.1063\pm0.0059, 4.9634\pm0.0028, 4.0978\pm0.0023, 4.5308\pm0.0020, 4.755\}\times10^{-4}$\ traps pixel$^{-1}$ day$^{-1}$. 
Dates at which the gradient changes were assessed by eye from \autoref{fig:density_evol} and held fixed while fitting other parameters. The final gradient is set to extrapolate the best-fitting single gradient (\autoref{eqn:rhotot_linear}) into the future.

\section{Results of CTI Correction}
\label{sec:results}

We correct all images using \arctic, with all sets of model parameters: whether they were measured from individual epochs, or interpolated as a function of time between them.
(We also correct images using \arctic models with non-instantaneous charge capture or inhomogeneous release, and indeed find them statistically no better than the fast model).
We correct the same images used for calibration but, because the final correction uses only 8 numbers fitted from nearly 200 sets of images, the contribution of any one image to those numbers is tiny: the correction of each image is thus almost independent of the model calibration.
To quantitatively assess the quality of correction, we repeat the process of \S\ref{sec:warm_pixel_measure}: re-identifying warm pixels, generating new trail profiles $\nelec^\mathrm{trail}(\Delta y;\nelec^{\rm wp},y)$ (an example of which is shown as the black data points in \autoref{fig:stacked_trails}), then measuring the amplitude of these residual trails, $\rhotot^{\rm corr}$.
In this application to delta-function images, we do not apply read noise correction because it merely adds noise, and we are not interested in pixel covariance as we might be for extended sources.

\subsection{Verification \& validation tests}
\label{sec:trail_fits:otf_validation}

Our previous algorithm to measure the amplitude of trails (\S\ref{sec:trail_fits}) is ill-suited to measuring the low or slightly negative (in the case of overcorrection) residual trails after correction, because \arctic requires positive-definite trap densities. We therefore fit residual trails using a sum-of-exponentials model 
\begin{eqnarray}
  \nelec(\Delta y) = \rhotot^{\rm corr} \left[
    \left(\dfrac{\nelec^{\rm wp} - d}{w - d}\right)^\beta
    - \left(\dfrac{n_{\rm bg} - d}{w - d}\right)^\beta \right]
    ~y~~~~~~~~~~~~~~~~~~~~~~~~ \nonumber \\
 \times~\sum_i^{\mathrm{a,b,c}} \frac{\rhoi}{\rhotot}
      \left(1 - \exp\left(\dfrac{-1}{\taui}\right)\right)
      \exp\left(\dfrac{1 - \Delta y}{\taui}\right)
 \label{eqn:exponential_model_rhotot}
\end{eqnarray}
that is similar to \autorefb{eqn:exponential_model} but now has only one free parameter, $\rhotot^\mathrm{corr}$, 
which can be either positive or negative. We fix all other parameters $\{\rhoa/\rhotot, \rhob/\rhotot, \rhoc/\rhotot, \taua, \taub, \tauc, \beta, d\}$ to those measured before correction. As found in \S\ref{sec:trail_fits}, a sum-of-exponentials model fits $\rhotot$ with 5--7\% percent multiplicative bias compared to \arctic: but that is good {\it absolute} accuracy considering the trail amplitudes are low after correction.

An alternative validation test that we find useful on other science data, uses \autorefb{eqn:exponential_model_rhotot} to fit trailing of the sky background into the CCD overscan region. However, since \acswfc has many hot pixels of $\sim$$10,000$~electrons, measurements of the trail from those is significantly higher signal-to-noise than those from a sky background of $\sim$$100$~electrons.

A limitation of the trap density metric is that it only estimates the reduction in the trails that \arctic is designed to recreate. A more demanding metric to quantify extra residuals that our model does not describe, is the mean value of the 12 pixels behind warm pixels before or after CTI correction (the red or black data points in \autoref{fig:stacked_trails}). For each of the 50 trail profiles at each epoch, we calculate
\begin{equation}
  \label{mean trail height reduction}
  \mu_{\nelec^\mathrm{trail}} \equiv
  \frac{ \sum{ \nelec^\mathrm{trail} ~ \sigma_{\nelec}^{-2} } }
       { \sum{ \sigma_{\nelec}^{-2} } }~
\end{equation}
and an equivalent version $\mu_{n^\mathrm{trail}} ^\mathrm{corr}$ after CTI correction, where $\sigma_{\nelec}$ is the uncertainty on each pixel (defined in \S\ref{sec:warm_pixel_measure}), and thus the mean is inverse-variance weighted.

We rely most upon fractional residuals such as $\langle\rhotot^\mathrm{corr}/\rhotot^\mathrm{\meas}\rangle$, because these are generally quoted elsewhere in the literature. However, they are weighted towards performance early in the mission when $\rhotot^\mathrm{\meas}$ was low and CTI was not really a problem. Absolute residuals like $\langle\rhotot^\mathrm{corr}\rangle$ are roughly constant over time (\autoreffigs{fig:raw_time_evol_rho} and \ref{fig:density_evol}), and perhaps a better indication of overall lifetime performance, particularly when most damage has accumulated near the end.

\subsection{Achieved performance of CTI correction}

Before correction, the trap density increases from about zero at launch to 4~traps per pixel in 2025, with mean $\langle\rhotot^\mathrm{\meas}\rangle=2.070\pm0.091$~traps per pixel in our particular epochs. Before correction, the mean value of trails increases from about zero to 16~electrons per pixel, with $\langle \mu_{n^\mathrm{trail}}^\mathrm{\meas} \rangle=8.391\pm0.350$.
We assess the precision with which our \arctic models correct CTI trailing over the lifetime of \acswfc, by comparing those same metrics after correction, $\rhotot^\mathrm{corr}$ and $\mu_{n^\mathrm{trail}}^\mathrm{corr}$.

The performance of CTI correction is excellent, with $99.350\pm0.123$~per cent of trails removed (top row of \autoref{tab:results}) when all \arctic model parameters are fitted to the data being corrected. 
This success is particularly notable because it was not the case in previous work. Previous calibration schemes, which omitted the second term in \autorefb{eqn:pre-CTI_amplitude} or used \autorefb{eqn:exponential_model_rhotot} underestimated trap densities, which had to be rescaled by further, artificial iteration to force $\langle\rhotot^\mathrm{corr}\rangle$ to zero. Accurate calibration using the full version of \autorefb{eqn:pre-CTI_amplitude} to model warm pixels, and \arctic to trail them, means we no longer need that step and can trust direct CTI calibration as a distinct process from CTI correction.

Performance improves even further when model parameters concerning device physics are forced to constant values for the lifetime of ACS, and only the total amount of damage, $\rhotot$, is fitted to the data being corrected (rows 2 and 3 of \autoref{tab:results}). The mean value of $\rhotot^\mathrm{corr}$ moves closer to zero, and the scatter ($\sqrt{n_\mathrm{epoch}-1}=\sqrt{177}$ times the uncertainties) is dramatically smaller. As in \S\ref{sec:calibrate_time_evol_physical}, this extremely positive result demonstrates that the device model represents the CCDs well, with more data beating down measurement noise on physically meaningful parameters.

If the amount of radiation damage is interpolated between calibration epochs, as it would need to be when correcting other science data, the performance of CTI correction degrades slightly and gains a lot of scatter for crude time series models (rows 4 to 6). However, performance is improved back to being consistent with ideal performance ($99.493\pm0.225$~per cent of trails removed; \autoref{tab:results} bottom row) if the interpolation uses a good model. In that case, it appears that the combined statistical information content at all epochs has combined to be as good as or better than that in individual epochs.

The least constrained parameters are $d$ and $\tauc$. According to our model, the first of these describes the effective physical size $\fV(\nelec)$ of clouds of a few electrons (the quantum mechanical interpretation of this is slightly vague). Our use of science data as calibration data is direct, but poorly constrains the functional form of $\fV(\nelec)$ at low $\nelec$, because $\nelec$ is never much less than 100. 
Notably, mean CTI residuals $\langle\rhotot^\mathrm{corr}\rangle$ get closer to (and are consistent with) zero and change sign when we lower $d$ from 94.1~electrons to 51.1~electrons: perhaps an intermediate value of $d$ could be tuned such that the residual for this dataset is exactly zero.
Improving upon our current analysis would require measurements of trailing in long dark exposures, short flat fields, or trailing into overscan pixels \citep[such measurements with the {\sl WFC3} camera have indeed revealed peculiar behaviour of $\fV(\nelec)$ at $\nelec\sim22$~electrons, and suggest that charge clouds with few electrons may be preferentially exposed to fewer traps with long release times;][]{Anderson2024}. 
This is not essential for correcting science data, but might improve the correction of e.g.\ dark and bias frames.
Measurements of trails beyond the 12-pixel window of our data would also better constrain the release time of the slowest traps, $\tauc$. This is not essential for CTI correction in images of isolated sources, because of the (now convenient) degeneracy between $\rhoc$ and $\tauc$, but it might improve observations of extended, low surface-brightness features \cite[by e.g.\ ARRAKIHS][]{ARRAKIHS}.

There are uncorrected residuals at the level of 0.2~electrons per pixel or 2~per~cent of the trails. The $\mu_{n^\mathrm{trail}}^\mathrm{\meas}$ metric is (intentionally) constructed to quantify effects beyond our model, which our analysis is therefore unable to explain. Nonetheless, we can speculate that our $\mu_{n^\mathrm{trail}}^\mathrm{\meas}$ residuals may represent either

\begin{itemize}

\item imperfect background or bias subtraction. Background measurement is uncertain at this level (see \S\ref{sec:warm_pixel_measure}). Sky backgrounds ought to be symmetric about a warm pixel, because the shutter takes only milliseconds to open or close \citep{Gilliland2003}. However, the superbias is intentionally higher downstream of hot pixels in order to account for continued charge leakage during readout, and that leakage may not be \revone{temporally stable.

\item residuals of other electronic readout} effects, such as bias striping, bias shift, or crosstalk, which are not removed by subtracting the superbias \citep{Ryon2023}.

\item potential trail identification errors. The $\mu_{n^\mathrm{trail}}^\mathrm{corr}=0.205$, $\rhotot^\mathrm{corr}=-0.070$ residual in the epoch shown in \autoref{fig:stacked_trails} is dominated by a single pixel: the twelfth pixel in the trail in the top-right panel. Excursions in excess of statistical noise are common, and seem to be more common in the final pixel, but we have not ascertained why. (With all parameters free, the residual for that epoch becomes $\mu_{n^\mathrm{trail}}^\mathrm{corr}= -0.014$, $\rhotot^\mathrm{corr}=-0.050$ with $d=73.2$).

\end{itemize}

\subsection{Anomalies}

Four points in \autoref{fig:density_evol} differ significantly from adjacent epochs (in 2010, 2014, and two in 2017). After further inspection, these turn out to be measurements from the only epochs with exposure times ($\leqslant$30\,seconds) much shorter than all the rest ($\sim$500\,seconds). 
Such data were not intended to be included (see Section~\ref{sec:warm_pixel_finding}). 
If they are removed when calculating performance metrics (but retained for model calibration, i.e.\ the best-fit model parameters are unchanged), the goodness of fit $\chi^2_r$ is decreased by $\sim$1.4 for all four $\rhotot(t)$ models considered, and all the metrics are lowered by $\sim$1$\sigma$ (improving three of four metrics).

Nonetheless, these epochs are instructive, because they test the model fit in a regime far from the deep extragalactic imaging for which it has been calibrated. The biggest effect during calibration is that the exposure times are not long enough for the dark current to accumulate as warm pixels with $\nelec^\mathrm{wp,obs}\gresim 3500$~electrons. Thus for these epochs, no data exist in the three rightmost columns of \autoref{fig:stacked_trails}. Severely restricting the dynamic range over which the well-filling model (\autoref{eqn:cloud_volume}) is constrained, adds uncertainty to its parameters. 
Worse, the amplitude of trails in the seventh bin is biased low, because the distribution of $\nelec^\mathrm{wp,obs}$ is truncated part-way through the bin and therefore skewed low. This introduces spurious extra curvature to \autoref{fig:trail_checks}, biasing $d$ and $\beta$ to anomalously low values in fits when all model parameters are free (\autoref{fig:raw_time_evol_all} and top panel of \autoref{fig:raw_time_evol_rho}).
In future studies, we advise that calibration data should span the same range of $\nelec$ as the science data that are to be corrected (and perhaps no more). The fit could also be made resilient to this effect by fitting individual warm pixels, or at least binning them more finely.

The second consequence of short exposures is that the sky background is low, and traps in a warm pixel are pre-filled by less sacrificial charge. The ArCTIc model should accommodate this, if it is correct. Since it does not, either \autorefb{eqn:cloud_volume} is incorrect when extrapolated into a regime of low $\nelec$ where it is currently unconstrained, or both assumptions are incorrect: that CTI is volume-driven and capture is instantaneous. The former is more likely, because we see no evidence for the latter in well-constrained regimes. Since the inferred trap density is anomalously high when all other model parameters are fixed (bottom panel of \autoref{fig:raw_time_evol_rho}; the anomalies could equally have been too low), this suggests that the model’s volume of traps between the low sky background and $\nelec^\mathrm{wp,obs}$ is too small. The model could therefore be improved by allowing $V(\nelec)$ to continue decreasing monotonically as $\nelec$ drops below $d$, instead of being truncated at zero. (That a quantity referred to as the `volume' would become negative is unproblematic in this analysis, because only the positive volume difference between two values of $\nelec$ is ever used.) 
\cite{Clarke2025} suggested a possible functional form for additional low-$\nelec$ complexity in a CCD manufactured by the same company.

\section{Discussion}
\label{sec:discussion}

\subsection{Limitations and caveats}

Our best-fit model of CTI in \acswfc\ enables excellent correction, but the multi-step process by which it was calibrated is not perfect. The model uses 8 free parameters, fitted from 178 epochs of data like \autoref{fig:stacked_trails}. Exploring some of the 8 dimensions separately from others (\S\ref{sec:calibrate_time_evol_physical} then \S\ref{sec:calibrate_time_evol_radiation}) made calculations tractable, but did not fully take into account degeneracies between parameters. In the future, a more sophisticated method could simultaneously fit all data, for example with likelihood optimisation of a Bayesian Hierarchical Model via Expectation Propagation \citep[][Nightingale et al.\ in prep.]{epaawol}. This might also make it possible to add extra parameters or freedom in the model, for example with some physical quantities $\rhoi(t)$, $\taui(t,T)$, etc.\ varying over time $t$ or as a function of temperature $T$. That may be most relevant for detectors with significant numbers of traps created during manufacturing, which are not necessarily of the same species as those created in orbit.

Our calibration measures only the mean density of traps per pixel ($\rho_i$ for each trap species~$i$), and our correction assumes a uniform distribution of traps. This is ideal for science cases with statistical requirements across a large field of view: by allowing traps to capture and release fractional numbers of electrons, this design choice minimises Poisson discretisation noise due to the unknown or uncertain locations of traps in a real CCD, and marginalises over the quantum mechanical processes for stochastic capture and release. The output of this algorithm is indistinguishable from the average of many randomly initialised runs with an algorithm that monitors individual traps \citep{Massey+2010}. 
However, some columns in \acswfc\ do clearly contain more traps than others, and hence show stronger trailing (see \autoref{fig:example_image_zooms}), and it is possible that there are variations across the detector or between CCD quadrants, because of differential levels of radiation shielding. Our current implementation will overcorrect some features in an image, and undercorrect others. 

The \arctic algorithm to mimic the effects of CTI is also not perfect. We have explicitly verified that assumptions (which also allow \arctic\ to run faster) of instantaneous capture and homogeneous trap release times are appropriate for \acswfc\ (see \S\ref{sec:trail_fits}).
For the sake of computational speed, we also approximate most of the millions of pixel-to-pixel transfers that occur during CCD readout as having the same effect on a charge cloud (we use $E_{n=410}$, which uses five main passes; see \S\ref{sec:express}). 
We similarly approximate each pixel-to-pixel transfer as instantaneous, with charge capture and release during only the main dwell time; whereas pixel-to-pixel transfer in real hardware involves three or four phases of sub-pixel shifts, each with a brief dwell time. Sub-pixel shifts affect every pixel-to-pixel transfer in the same way, and \arctic\ has now been run with this complexity. In a case where the physical width of each phase is identical, we have verified that the effect of traps in extra phases or the barrier phases between pixels is equivalent to rescaling the trap density. Trap densities in this model are therefore not necessarily real, but {\it effective} densities that reproduce the same effect. We have not attempted to optimise for \acswfc the effect of some phases being narrower than others, and pushing a charge packet to higher regions of a (sub-)pixel, where it may encounter additional traps \citep{Hall2014,Hall+2014b} whose occupancy levels have not reached the same steady state.
In \arctic, this can be accommodated by adjusting parameters of the well-filling model (\S\ref{sec:arctic:ccd}) separately for each phase.

When correcting CTI, we assume that $n_\mathrm{iter}=5$ iterations is sufficient to converge. Further iterations change the image by $\ll$~1~electron, and by less than the read noise, but whether this is sufficient ought to depend on measurement requirements for a particular science goal. More profoundly, we also assume that trailing will never be more than a perturbation of the true image. This lets us initialise the iteration to correct CTI (\autoref{tab:remove_cti}) using the downloaded image itself. If enough radiation damage were to accrue, the mapping from noisy image A to true image $I$ would become many-to-one, and pixel-level CTI correction would become ambiguous. A workaround with more extreme radiation damage might be to forward-model image features: adding CTI to a noise-free model (e.g.\ of a galaxy, with morphology parameters informed by priors, or a modelled PSF: \citealt{Rowell2021}) then matching it to data, rather than trying to directly remove CTI from noisy data.

\subsection{Future improvements}

The next thing to improve about \arctic is probably the well-filling model (\autoref{eqn:cloud_volume}) at low $\nelec$. The discontinuity in $\mathrm{d}\fV/\mathrm{d}\nelec$ at $\nelec=d$ creates a threshold below which CTI is absent (because $\mathrm{d}\fV/\mathrm{d}\nelec$=$0$), and above which CTI acts with greatest effect (because $\mathrm{d}\fV/\mathrm{d}\nelec$ is maximal). 
This primarily affects the correction of images with low signal $\nelec$$\sim$$d$, such as bias frames. 
The gradient discontinuity does not statistically bias such data, but it has an asymmetric effect on positive or negative noise peaks that emerges in the pixel covariance matrix. 
A more physically complete model of device physics could add low-$\nelec$ complexity to \autorefb{eqn:cloud_volume}; and a different implementation could symmetrise $\fV(\nelec)$ about $\nelec=d$, or at least avoid zero gradient.
To calibrate \arctic\ parameters in a way that is resilient to extreme values of $\nelec$, future fits should ideally be performed simultaneously on all trails, or at least in finer bins of $\nelec^\mathrm{wp,obs}$ and distance from readout register.
If the faintest CTI trails were then seen to be steeper than others \citep[c.f.][]{Anderson2024}, it would also be feasible to tie different values of $\beta$ to different trap species.

Another asymmetry in \arctic is that we require a positive density of charge traps $\rho_{i}>0$. It is worth considering whether to allow trap species with negative density, which capture negative numbers of electrons (i.e.\ release electrons). This would simplify calibration of residual trails close to zero (see \S\ref{sec:trail_fits:otf_validation}). However, simply flipping the sign of trap densities could still not be used for more than the first iteration of CTI correction (\autoref{tab:remove_cti}). It cannot be extended into an iteration because the number of traps filled by electrons during trailing of image $I$ is not the same as the number of traps filled during trailing of image $A$ --- which is why removing CTI is not the same as adding negative CTI.

It is possible to quickly process {\it spatial} windows of readout, by setting to zero some elements of the express matrix (\S\ref{sec:express}). It would also be desirable to be able to process {\it temporal} windows of readout, for example to add or correct cosmic rays that hit during readout, and are therefore less trailed than the pixels around them. 

CTI correction of both {\sl HST}/{\sl ACS} and Gaia is simplified by the dominance of CTI in one readout direction, \citep[although cross-CTI has recently been detected in ACS by][]{Ryon2024}. This should be included in future measurements of ACS CTI. If \arctic were further extended, several other electronic readout effects, such as blooming (not bleeding) or pixel bounce \citep{Janesick2000}, could also be efficiently corrected at the same time as CTI.

\section{Conclusions}
\label{sec:conclusions}

The longevity of the {\sl Hubble Space Telescope}'s {\sl Advanced Camera for Surveys} in low Earth orbit has enabled almost continuous monitoring of radiation damage with a single, highly sensitive detector for 23~years. Warm pixels in the camera ought to appear as $\delta$-functions, and deviations from this are an effective way to measure damage. We find that the rate of damage is modulated by $18.5^{+4.5}_{-0.5}$~per cent over two 11~year Solar cycles, peaking $430^{+11}_{-5}$~days after Solar minimum. We interpret this as being proportional to the high-energy radiation flux from Solar protons and galactic cosmic rays. The main departure from this is a recent, unexplained reduction in the rate of damage since 2024.

Detailed properties of the damage to \acswfc are consistent with dislocations to the silicon lattice in three dominant topological configurations. These `traps' capture nearby free/valence electrons in a time that is consistent with zero, or at least $\ll t_\mathrm{dwell}=3212\,\muup$s, the parallel clock time. The spread of their characteristic release times is also consistent with zero: traps of one species all appear alike, as if they have \revone{settled into low-energy states}.

That our models of both the radiation environment and the detector damage are accurate (or at least degenerate with each other) is supported by their use in combination to successfully remove spurious trailing due to Charge Transfer Inefficiency in science imaging. 
That the model parameters are physically meaningful (rather than just convenient parametric forms) is supported by the improvement in correction accuracy at all epochs, when noise is beaten down by combining measurements from different epochs. There are still $\sim0.2$~electrons/pixel residuals that the \arctic model of parallel CTI does not capture. These may be due to mismodelling of bias or background during calibration, 
or other electronic detector effects (which may or may not be a problem for different science applications).

Removing 99.9\% of CTI trailing in the most recent \acswfc data (99.5\% of trailing averaged over the camera lifetime) passes a significant milestone: for the first time, this meets requirements for the ESA Euclid mission \citep{Israel+2015,Paykari2020}. Those requirements have relaxed slightly because (perhaps like our measurements in LEO) the current radiation environment at the Earth-Sun Lagrange point L2 is not as hostile as predicted by the SPENVIS model \citep{Skottfelt2024}.
Importantly for future models of CTI, we have now understood which approximations can be made during CTI calibration to still achieve this accuracy of CTI correction --- and which approximations introduce unacceptable model biases. 
In particular, accounting for uncertainty in background subtraction prevents the need for {\it ad~hoc} upweighting of data at greater distances from the readout register; and adding the few trailed electrons back to a pre-CTI model of each warm pixel (\S\ref{sec:warm_pixel_measure}) removes the dominant calibration bias.
This neatly separates the tasks of CTI calibration and CTI correction, so that each can be performed without feedback from the other (although we have also proposed an on-the-fly validation \& verification test that can check the success of CTI correction at low signal-to-noise on any science image).

\vspace{-4mm}
\section*{Acknowledgements}
We thank Claudio Pagani, Alastair Edge, Jay Anderson, Jenna Ryon, Mark Swinbank, David Harvey and the referee for helpful feedback. 
Authors were supported in Durham by the UK STFC (grant ST/X001075/1) and the UK Space Agency (grant ST/X001997/1).
JAK and JWN are also now supported by STFC Ernest Rutherford Fellowships.
ZDL is supported by STFC studentship ST/Y509346/1.
Our statistical analysis software was partially developed through STFC grant ST/T002565/1, then  InnovateUK grants TS/V002856/1 and TS/Y014693/1.
This work used the DiRAC@Durham facility managed by the Institute for Computational Cosmology on behalf of the STFC DiRAC HPC facility (www.dirac.ac.uk). The equipment was funded by BEIS capital funding via STFC capital grants ST/K00042X/1, ST/P002293/1, ST/R002371/1 and ST/S002502/1, Durham University and STFC operations grant ST/R000832/1. DiRAC is part of the UK National e-Infrastructure.

\vspace{-3mm}
\section*{Data Availability}
The data presented in this paper were obtained from the Mikulski Archive for Space Telescopes (MAST). STScI is operated by the Association of Universities for Research in Astronomy, Inc., under NASA contract NAS5-26555. Support for MAST for non-HST data is provided by the NASA Office of Space Science via grant NNX13AC07G and by other grants and contracts. 
\vspace{-3mm}

\bibliographystyle{mnras}
\bibliography{cti.bib}

@article{Murray2013,
author = {Murray, Neil J and Burt, David J and Holland, Andrew D and Stefanov, Konstantin D and Gow, Jason P D and MacCormick, Calum and Dryer, Ben J and Allanwood, Edgar A H},
doi = {10.1117/12.2024839},
isbn = {9780819497109},
issn = {0277786X},
journal = {UV/Optical/IR Space Telescopes and Instruments: Innovative Technologies and Concepts VI},
pages = {88600K},
title = {{Multi-level parallel clocking of CCDs for: improving charge transfer efficiency, clearing persistence, clocked anti-blooming, and generating low-noise backgrounds for pumping}},
url = {http://proceedings.spiedigitallibrary.org/proceeding.aspx?doi=10.1117/12.2024839},
volume = {8860},
year = {2013}
}

@article{Hall2014,
author = {Hall, David J and Murray, Neil J and Holland, Andrew D and Gow, Jason and Clarke, Andrew and Burt, David},
doi = {10.1109/TNS.2013.2295941},
isbn = {9781467350570},
issn = {00189499},
journal = {IEEE Transactions on Nuclear Science},
number = {4},
pages = {1826--1833},
title = {{Determination of in situ trap properties in CCDs using a "single-trap pumping" technique}},
volume = {61},
year = {2014}
}

@article{Gow2016,
author = {Gow, Jason P. D. and Murray, Neil J.},
doi = {10.1117/12.2232706},
isbn = {9781510602090},
issn = {1996756X},
journal = {High Energy, Optical, and Infrared Detectors for Astronomy VII},
number = {0},
pages = {99152A},
title = {{Simplified charge transfer inefficiency correction in CCDs by trap-pumping}},
url = {http://proceedings.spiedigitallibrary.org/proceeding.aspx?doi=10.1117/12.2232706},
volume = {9915},
year = {2016}
}

@article{Anderson2010,
archivePrefix = {arXiv},
arxivId = {1007.3987},
author = {Anderson, Jay and Bedin, Luigi R.},
doi = {10.1086/656399},
eprint = {1007.3987},
issn = {0004-6280},
journal = {Publ. Astron. Soc. Pac.},
month = {sep},
number = {895},
pages = {1035--1064},
title = {{An Empirical Pixel-Based Correction for Imperfect CTE. I. HST 's Advanced Camera for Surveys1}},
url = {http://iopscience.iop.org/article/10.1086/656399},
volume = {122},
year = {2010}
}

@article{Rhodes2007,
archivePrefix = {arXiv},
arxivId = {astro-ph/0702140},
author = {Rhodes, Jason D. and Massey, Richard J. and Albert, Justin and Collins, Nicholas and Ellis, Richard S. and Heymans, Catherine and Gardner, Jonathan P. and Kneib, Jean‐Paul and Koekemoer, Anton and Leauthaud, Alexie and Mellier, Yannick and Refregier, Alexander and Taylor, James E. and {Van Waerbeke}, Ludovic},
doi = {10.1086/516592},
eprint = {0702140},
issn = {0067-0049},
journal = {The Astrophysical Journal Supplement Series},
month = {sep},
number = {1},
pages = {203--218},
primaryClass = {astro-ph},
title = {{                    The Stability of the Point‐Spread Function of the Advanced Camera for Surveys on the                    Hubble Space Telescope                    and Implications for Weak Gravitational Lensing                  }},
url = {http://stacks.iop.org/0067-0049/172/i=1/a=203},
volume = {172},
year = {2007}
}

@article{Short2013,
archivePrefix = {arXiv},
arxivId = {1302.1416},
author = {Short, A. and Crowley, C. and de Bruijne, J. H.J. and Prod'homm, T.},
doi = {10.1093/mnras/stt114},
eprint = {1302.1416},
issn = {00358711},
journal = {MNRAS},
number = {4},
pages = {3078--3085},
title = {{An analytical model of radiation-induced charge transfer inefficiency for CCD detectors}},
volume = {430},
year = {2013}
}

@article{Bristow2003,
author = {Bristow, Paul},
journal = {preprint (astro-ph/0310714)},
month = {oct},
title = {{Application of Model Derived Charge Transfer Inefficiency Corrections to STIS Photometric CCD Data}},
year = {2003}
}

@article{Speagle2020,
archivePrefix = {arXiv},
arxivId = {1904.02180},
author = {Speagle, Joshua S},
doi = {10.1093/mnras/staa278},
eprint = {1904.02180},
issn = {0035-8711},
journal = {MNRAS},
number = {3},
pages = {3132--3158},
title = {{dynesty: a dynamic nested sampling package for estimating Bayesian posteriors and evidences}},
volume = {493},
year = {2020}
}

@article{Skilling2006,
archivePrefix = {arXiv},
arxivId = {arXiv:1011.1669v3},
author = {Skilling, John},
doi = {10.1214/06-BA127},
eprint = {arXiv:1011.1669v3},
isbn = {9788578110796},
issn = {19360975},
journal = {Bayesian Analysis},
number = {4},
pages = {833--860},
pmid = {25246403},
title = {{Nested sampling for general Bayesian computation}},
url = {http://projecteuclid.org/euclid.ba/1340370944},
volume = {1},
year = {2006}
}

@article{Paykari2020,
archivePrefix = {arXiv},
arxivId = {astro-ph.CO/1910.10521},
author = {{Euclid Collaboration} and Paykari, P and Kitching, T and Hoekstra, H and Azzollini, R and Cardone, V.$\sim$F. and Cropper, M and Duncan, C.$\sim$A.$\sim$J. and Kannawadi, A and Miller, L and Aussel, H and Conti, I.$\sim$F. and Auricchio, N and Baldi, M and Bardelli, S and Biviano, A and Bonino, D and Borsato, E and Bozzo, E and Branchini, E and Brau-Nogue, S and Brescia, M and Brinchmann, J and Burigana, C and Camera, S and Capobianco, V and Carbone, C and Carretero, J and Castander, F.$\sim$J. and Castellano, M and Cavuoti, S and Charles, Y and Cledassou, R and Colodro-Conde, C and Congedo, G and Conselice, C and Conversi, L and Copin, Y and Coupon, J and Courtois, H.$\sim$M. and {Da Silva}, A and Dupac, X and Fabbian, G and Farrens, S and Ferreira, P.$\sim$G. and Fosalba, P and Fourmanoit, N and Frailis, M and Fumana, M and Galeotta, S and Garilli, B and Gillard, W and Gillis, B.$\sim$R. and Giocoli, C and Graci{\'{a}}-Carpio, J and Grupp, F and Hormuth, F and Ili{\'{c}}, S and Israel, H and Jahnke, K and Keihanen, E and Kermiche, S and Kilbinger, M and Kirkpatrick, C.$\sim$C. and Kubik, B and Kunz, M and Kurki-Suonio, H and Laureijs, R and {Le Mignant}, D and Ligori, S and Lilje, P.$\sim$B. and Lloro, I and Maciaszek, T and Maiorano, E and Marggraf, O and Markovic, K and Martinet, N and Marulli, F and Massey, R and Mauri, N and Medinaceli, E and Mei, S and Mellier, Y and Meneghetti, M and Metcalf, R.$\sim$B. and Moresco, M and Moscardini, L and Munari, E and Neissner, C and Nichol, R.$\sim$C. and Niemi, S and Nutma, T and Padilla, C and Paltani, S and Pasian, F and Pettorino, V and Pires, S and Polenta, G and Raison, F and Renzi, A and Rhodes, J and Romelli, E and Roncarelli, M and Rossetti, E and Saglia, R and Sakr, Z and S{\'{a}}nchez, A.$\sim$G. and Sapone, D and Scaramella, R and Schneider, P and Schrabback, T and Scottez, V and Secroun, A and Serrano, S and Sirignano, C and Sirri, G and Stanco, L and Starck, J -L. and Sureau, F and Tallada-Cresp\'\i, P and Taylor, A and Tenti, M and Tereno, I and Toledo-Moreo, R and Torradeflot, F and Valenziano, L and Vannier, M and Vassallo, T and Zoubian, J and Zucca, E},
doi = {10.1051/0004-6361/201936980},
eprint = {1910.10521},
journal = {A\&A},
month = {mar},
pages = {A139},
primaryClass = {astro-ph.CO},
title = {{Euclid preparation. VI. Verifying the performance of cosmic shear experiments}},
volume = {635},
year = {2020}
}

@article{Cropper2013,
author = {Cropper, Mark and Hoekstra, Henk and Kitching, Thomas and Massey, Richard and Amiaux, J{\'{e}}r{\^{o}}me and Miller, Lance and Mellier, Yannick and Rhodes, Jason and Rowe, Barnaby and Pires, Sandrine and Saxton, Curtis and Scaramella, Roberto},
doi = {10.1093/mnras/stt384},
issn = {00358711},
journal = {MNRAS},
number = {4},
pages = {3103--3126},
title = {{Defining a weak lensing experiment in space}},
volume = {431},
year = {2013}
}

@ARTICLE{sidc,
   author = {{SILSO World Data Center}},
  address = {Royal Observatory of Belgium, avenue Circulaire 3, 1180 Brussels, Belgium},
    title = "{The International Sunspot Number}",
  journal = {International Sunspot Number Monthly Bulletin and online catalogue},
     year = "2000-2024",
    month = {},
   volume = {},
    pages = {},
   adsurl = {https://www.sidc.be/SILSO/datafiles}
}

@MISC{Biretta+Kozhurina-Platais2005,
    author = {{Biretta}, J. and {Kozhurina-Platais}, V.},
    title = "{Hot Pixels as a Probe of WFPC2 CTE Effects}",
    howpublished = {Instrument Science Report WFPC2 2005-01},
    year = 2005,
    month = jul,
    pages = {23},
    adsurl = {https://ui.adsabs.harvard.edu/abs/2005wfpc.rept....1B}
}

@MISC{Coe+Grogin2014,
    author = {{Coe}, D. and {Grogin}, N.},
    title = "{Readout Dark: Dark Current Accumulation During ACS/WFC Readout}",
    howpublished = {Instrument Science Report ACS 2014-02},
    year = 2014,
    month = dec,
    pages = {17},
    adsurl = {https://ui.adsabs.harvard.edu/abs/2014acs..rept....2C}
}

@ARTICLE{Hall+2014b,
    author = {{Hall}, David J. and {Murray}, Neil J. and {Holland}, Andrew D. and
    {Gow}, Jason and {Clarke}, Andrew and {Burt}, David},
    title = "{Determination of In Situ Trap Properties in CCDs Using a ``Single-Trap Pumping'' Technique}",
    journal = {IEEE Transactions on Nuclear Science},
    year = "2014",
    month = "Aug",
    volume = {61},
    number = {4},
    pages = {1826-1833},
    doi = {10.1109/TNS.2013.2295941},
    adsurl = {https://ui.adsabs.harvard.edu/abs/2014ITNS...61.1826H}
}

@ARTICLE{Israel+2015,
    author = {{Israel}, Holger and {Massey}, Richard and {Prod'homme}, Thibaut and
    {Cropper}, Mark and {Cordes}, Oliver and {Gow}, Jason and
    {Kohley}, Ralf and {Marggraf}, Ole and {Niemi}, Sami and
    {Rhodes}, Jason and {Short}, Alex and {Verhoeve}, Peter},
    title = "{How well can charge transfer inefficiency be corrected? A parameter sensitivity study for iterative correction}",
    journal = {\mnras},
    year = "2015",
    month = "Oct",
    volume = {453},
    number = {1},
    pages = {561-580},
    doi = {10.1093/mnras/stv1660},
    archivePrefix = {arXiv},
    eprint = {1506.07831},
    primaryClass = {astro-ph.IM},
    adsurl = {https://ui.adsabs.harvard.edu/abs/2015MNRAS.453..561I}
}

@article{Lindegren1998,
  title={Charge trapping effects in ccds for gaia astrometry},
  author={Lindegren, L},
  journal={internal Gaia Livelink, SAG-LL-022},
  year={1998}
}

@ARTICLE{Massey2010,
    author = {{Massey}, Richard},
    title = "{Charge transfer inefficiency in the Hubble Space Telescope since Servicing Mission 4}",
    journal = {\mnras},
    year = "2010",
    month = "Nov",
    volume = {409},
    number = {1},
    pages = {L109-L113},
    doi = {10.1111/j.1745-3933.2010.00959.x},
    archivePrefix = {arXiv},
    eprint = {1009.4335},
    primaryClass = {astro-ph.IM},
    adsurl = {https://ui.adsabs.harvard.edu/abs/2010MNRAS.409L.109M}
}

@article{Massey+2010,
    author = {Massey, Richard and Stoughton, Chris and Leauthaud, Alexie and Rhodes, Jason and Koekemoer, Anton and Ellis, Richard and Shaghoulian, Edgar},
    title = "{Pixel-based correction for Charge Transfer Inefficiency in the Hubble Space Telescope Advanced Camera for Surveys}",
    journal = {\mnras},
    volume = {401},
    number = {1},
    pages = {371-384},
    year = {2009},
    month = {12},
    issn = {0035-8711},
    doi = {10.1111/j.1365-2966.2009.15638.x},
    url = {https://doi.org/10.1111/j.1365-2966.2009.15638.x},
    eprint = {https://academic.oup.com/mnras/article-pdf/401/1/371/18581537/mnras0401-0371.pdf},
}

@ARTICLE{Massey+2014,
    author = {{Massey}, Richard and {Schrabback}, Tim and {Cordes}, Oliver and
    {Marggraf}, Ole and {Israel}, Holger and {Miller}, Lance and
    {Hall}, David and {Cropper}, Mark and {Prod'homme}, Thibaut and
    {Niemi}, Sami-Matias},
    title = "{An improved model of charge transfer inefficiency and correction algorithm for the Hubble Space Telescope}",
    journal = {\mnras},
    year = "2014",
    month = "Mar",
    volume = {439},
    number = {1},
    pages = {887-907},
    doi = {10.1093/mnras/stu012},
    archivePrefix = {arXiv},
    eprint = {1401.1151},
    primaryClass = {astro-ph.IM},
    adsurl = {https://ui.adsabs.harvard.edu/abs/2014MNRAS.439..887M}
}

@MISC{Ogaz+2013,
    author = {{Ogaz}, S. and {Anderson}, J. and {Maybhate}, A. and {Smith}, L.},
    title = "{Column Dependency in Charge Transfer Efficiency Correction}",
    howpublished = {ACS Instrument Science Report 2013-02},
    year = 2013,
    month = jul,
    pages = {2},
    adsurl = {https://ui.adsabs.harvard.edu/abs/2013acs..rept....2O}
}

@MISC{Ryon2024,
    author = {{Ryon}, J. and {Grogin}, N.},
    title = "{Serial Charge Transfer Efficiency in ACS/WFC}",
    howpublished = {ACS Instrument Science Report 2024-07},
    year = 2024,
    month = December,
    volume = {2024-07},
    number = {07},
    adsurl = {https://www.stsci.edu/hst/instrumentation/acs/documentation/instrument-science-reports-isrs},
}

@MISC{Guzman2024,
    author = {{Guzman}, A. and {Ryon}, J.},
    title = "{Evolution of Sink Pixels in ACS/WFC and Connection to Charge Transfer Efficiency}",
    howpublished = {ACS Instrument Science Report 2024-01},
    year = 2024,
    month = April,
    volume = {2024-01},
    number = {01},
    adsurl = {https://www.stsci.edu/hst/instrumentation/acs/documentation/instrument-science-reports-isrs},
}

@ARTICLE{Rhodes+2010,
    author = {{Rhodes}, Jason and {Leauthaud}, Alexie and {Stoughton}, Chris and
    {Massey}, Richard and {Dawson}, Kyle and {Kolbe}, William and
    {Roe}, Natalie},
    title = "{The Effects of Charge Transfer Inefficiency (CTI) on Galaxy Shape Measurements}",
    journal = {\pasp},
    year = "2010",
    month = "Apr",
    volume = {122},
    number = {890},
    pages = {439},
    doi = {10.1086/651675},
    archivePrefix = {arXiv},
    eprint = {1002.1479},
    primaryClass = {astro-ph.IM},
    adsurl = {https://ui.adsabs.harvard.edu/abs/2010PASP..122..439R}
}

@INCOLLECTION{Ryon2023,
       author = {{Ryon}, J.~E.},
        title = "{ACS Instrument Handbook for Cycle 31 v. 22.0}",
    booktitle = {ACS Instrument Handbook for Cycle 31 v. 22.0},
         year = 2023,
       adsurl = {https://ui.adsabs.harvard.edu/abs/2023acsi.book...22R},
    publisher = {STScI}
}

@ARTICLE{ShockleyRead1952,
    author = {{Shockley}, W. and {Read}, W.~T.},
    title = "{Statistics of the Recombinations of Holes and Electrons}",
    journal = {Physical Review},
    year = "1952",
    month = "Sep",
    volume = {87},
    number = {5},
    pages = {835-842},
    doi = {10.1103/PhysRev.87.835},
    adsurl = {https://ui.adsabs.harvard.edu/abs/1952PhRv...87..835S}
}

@inproceedings{Hall1951,
  title={Germanium rectifier characteristics},
  author={Hall, RN},
  booktitle={Physical Review},
  volume={83},
  number={1},
  pages={228--228},
  year={1951},
  organization={AMERICAN PHYSICAL SOC ONE PHYSICS ELLIPSE, COLLEGE PK, MD 20740-3844 USA}
}

@MISC{Sirianni+2006,
    author = {{Sirianni}, M. and {Gilliland}, R. and {Sembach}, K.},
    title = "{The ACS Side-2 Switch: An Opportunity to Adjust the WFC CCD Temperature Setpoint}",
    howpublished = {Instrument Science Report ACS 2006-002},
    year = 2006,
    month = sep
}

@MISC{Mack+2007,
    author = {{Mack}, J. and {Gilliland}, R. and {Anderson}, J. and {Sirianni}, M. },
    title = "{The ACS Side-2 Switch: An Opportunity to Adjust the WFC CCD Temperature Setpoint}",
    howpublished = {Instrument Science Report ACS 2007-002},
    year = 2007
}

@MISC{Hopkinson2001,
    author = {{Hopkinson}, G. },
    journal = "IEEE Trans.\ Nuclear Sci.",
    year = "2001",
    volume = {48},
    pages = {6}
}

@article{Janesick1987,
  title={Scientific charge-coupled devices},
  author={Janesick, James R and Elliott, Tom and Collins, Stewart and Blouke, Morley M and Freeman, Jack},
  journal={Optical Engineering},
  volume={26},
  number={8},
  pages={692--714},
  year={1987},
  publisher={SPIE}
}

@incollection{Jerram2020,
  title={CMOS and CCD image sensors for space applications},
  author={Jerram, P and Stefanov, K},
  booktitle={High performance silicon imaging},
  pages={255--287},
  year={2020},
  publisher={Elsevier}
}

@article{Dawson2008,
  title={Radiation tolerance of fully-depleted P-channel CCDs designed for the SNAP satellite},
  author={Dawson, Kyle and Bebek, Chris and Emes, John and Holland, Steve and Jelinsky, Sharon and Karcher, Armin and Kolbe, William and Palaio, Nick and Roe, Natalie and Saha, Juhi and others},
  journal={IEEE Transactions on Nuclear Science},
  volume={55},
  number={3},
  pages={1725--1735},
  year={2008},
  publisher={IEEE}
}

@article{Skottfelt2024,
  title={Tracking radiation damage of Euclid VIS detectors after 1 year in space},
  author={Skottfelt, Jesper and Wander, Matt and Cropper, Mark and Dryer, Ben and Hall, David J and Hayes, Richard and Kelman, Bradley and Kitching, Tom and Kohley, Ralf and Lagattuta, David and others},
  journal={arXiv preprint arXiv:2407.01268},
  year={2024}
}

@inproceedings{Ahmed2020,
  title={Gaia CCDs: charge transfer inefficiency measurements between five years of flight},
  author={Ahmed, Saad and Hall, David and Crowley, Cian and Skottfelt, Jesper and Dryer, Ben and Seabroke, George and Hernandez, Jose and Holland, Andrew},
  booktitle={X-Ray, Optical, and Infrared Detectors for Astronomy IX},
  volume={11454},
  pages={155--166},
  year={2020},
  organization={SPIE}
}

@MISC{Bushouse2011,
       author = {{Bushouse}, H. and {Baggett}, S. and {Gilliland}, R. and {Noeske}, K. and {Petro}, L.},
        title = "{WFC3/UVIS Charge Injection Behavior: Results of an Initial Test}",
 howpublished = {WFC3 Instrument Science Report 2011-02, 11 pages},
         year = 2011,
        month = jan,
        pages = {2},
       adsurl = {https://ui.adsabs.harvard.edu/abs/2011wfc..rept....2B}
}

@PHDTHESIS{Bilgi2019,
       author = {{Bilgi}, Pavaman},
        title = "{Optimization of CCD charge transfer for ground and space-based astronomy}",
       school = {California Institute of Technology},
         year = 2019,
        month = jan,
       adsurl = {https://ui.adsabs.harvard.edu/abs/2019PhDT.......198B}
}

@article{Rivers2023,
	author = {Rivers, Xavier and Ryon, Jenna and Cohen, Yotam},
	journal = {Bulletin of the AAS},
	number = {2},
	year = {2023},
	month = {jan 31},
	note = {https://baas.aas.org/pub/2023n2i462p07},
	publisher = {},
	title = {Characterizing {Serial} {CTE} ({XCTE}) in {Hubble}\textquoteright{}s {ACS}/{WFC} {Detector}},
	volume = {55},
}

@article{Astier2023,
	author = {{Astier}, Pierre and {Regnault}, Nicolas},
	doi = {10.1051/0004-6361/202245407},
	journal = {A&A},
	pages = {A118},
	title = {Correction of the brighter-fatter effect on the CCDs of Hyper Suprime-Cam},
	url = {https://doi.org/10.1051/0004-6361/202245407},
	volume = 670,
	year = 2023}

@article{Ali2022,
	article-number = {9341},
	author = {Ali, Munir and Dong, Yunfan and Lv, Jianhang and Guo, Hongwei and Abid Anwar, Muhammad and Tian, Feng and Shahzad, Khurram and Liu, Wei and Yu, Bin and Bodepudi, Srikrishna Chanakya and Xu, Yang},
	doi = {10.3390/s22239341},
	issn = {1424-8220},
	journal = {Sensors},
	number = {23},
	pubmedid = {36502042},
	title = {In-Situ Monitoring of Reciprocal Charge Transfer and Losses in Graphene-Silicon CCD Pixels},
	url = {https://www.mdpi.com/1424-8220/22/23/9341},
	volume = {22},
	year = {2022}}

@ARTICLE{Kelman2023,
       author = {{Kelman}, Bradley},
        title = "{Investigation of an irradiated CCD device: building and testing a Charge Transfer Inefficiency correction pipeline using the Pyxel framework}",
      journal = {PSD13: The 13th International Conference on Position Sensitive Detectors},
         year = 2023,
        month = September,
       volume = {64},
       number = {13},
       adsurl = {https://indico.cern.ch/event/1230837/contributions/5518061/},
 }

@MISC{Desjardins2019,
    author = {{Desjardins}, T.~D.},
    title = "{Post-SM4 ACS/WFC Bias I: The Read Noise History}",
    howpublished = {ACS Instrument Science Report ACS 2019-02},
    year = 2019,
    month = mar,
    volume = {2019-02},
    number = {02},
    adsurl = {https://ui.adsabs.harvard.edu/abs/2019acs..rept....2D}
}

@MISC{Ryon2022,
    author = {{Ryon}, J. and {Grogin}, N. and {McDonald}, M.},
    title = "{Fading Hot Pixels in ACS/WFC}",
    howpublished = {ACS Instrument Science Report ACS 2022-07},
    year = 2022,
    month = dec,
    volume = {2022-07},
    number = {07}
}

@MISC{Gilliland2003,
    author = {{Gilliland}, R. and {Hartig}, G.},
    title = "{Stability and Accuracy of HRC and WFC Shutters}",
    howpublished = {ACS Instrument Science Report ACS 2003-03},
    year = 2003,
    month = jun,
    volume = {2003-03},
    number = {03}
}

@ARTICLE{Anand2023,
       author = {{Anand}, G. and {Grogin}, N. and {Anderson}, J. and {Ryon}, J.},
        title = "{Systematic Effects in ACS/WFC Absolute Gain Measurements}",
      journal = {HST ISR},
         year = 2023,
        month = April,
       volume = {2023-02},
       number = {02},
       adsurl = {https://www.stsci.edu/hst/instrumentation/acs/documentation/instrument-science-reports-isrs},
 }

@ARTICLE{Stark2024,
       author = {{Stark}, David and {Grogin}, Norman},
        title = "{The Impact of CTE on Point Source Detection in Simulated ACS/WFC Imaging Data}",
      journal = {HST ISR},
         year = 2024,
        month = May,
       volume = {2024-02},
       number = {02},
       adsurl = {https://www.stsci.edu/hst/instrumentation/acs/documentation/instrument-science-reports-isrs},
 }

@ARTICLE{Chiaberge2022,
       author = {{Chiaberge}, M. and {Ryon}, J.},
        title = "{ACS/WFC CTE photometric correction: improved model for bright point sources}",
      journal = {HST ISR},
         year = 2022,
        month = December,
       volume = {2022-06},
       number = {06},
       adsurl = {https://www.stsci.edu/hst/instrumentation/acs/documentation/instrument-science-reports-isrs},
 }

@ARTICLE{Anderson2024,
       author = {{Anderson}, J.},
        title = "{Revisiting x-CTE in WFC3/UVIS}",
      journal = {HST ISR},
         year = 2024,
        month = June,
       volume = {2024-07},
       number = {07},
       adsurl = {https://www.stsci.edu/files/live/sites/www/files/home/hst/instrumentation/wfc3/documentation/instrument-science-reports-isrs/_documents/2024/WFC3-ISR-2024-07.pdf},
 }

@ARTICLE{Anderson2018,
       author = {{Anderson}, J. and {Ryon}, J.},
        title = "{Improving the Pixel-Based CTE-correction Model for ACS/WFC}",
      journal = {HST ISR},
         year = 2018,
        month = August,
       volume = {2018-04},
       number = {04},
       adsurl = {https://www.stsci.edu/hst/instrumentation/acs/documentation/instrument-science-reports-isrs},
 }

@ARTICLE{Zhao2021,
       author = {{Zhao}, F. and {Lo Curto}, G. and {Pasquini}, L. and {Gonz{\'a}lez Hern{\'a}ndez}, J.~I. and {De Medeiros}, J.~R. and {Canto Martins}, B.~L. and {Le{\~a}o}, I.~C. and {Rebolo}, R. and {Su{\'a}rez Mascare{\~n}o}, A. and {Esposito}, M. and {Manescau}, A. and {Steinmetz}, T. and {Udem}, T. and {Probst}, R. and {Holzwarth}, R. and {Zhao}, G.},
        title = "{Measuring and characterizing the line profile of HARPS with a laser frequency comb}",
      journal = {\aap},
         year = 2021,
        month = jan,
       volume = {645},
          eid = {A23},
        pages = {A23},
          doi = {10.1051/0004-6361/201937370},
archivePrefix = {arXiv},
       eprint = {2011.03391},
 primaryClass = {astro-ph.IM},
       adsurl = {https://ui.adsabs.harvard.edu/abs/2021A&A...645A..23Z}
}

@inproceedings{Skottfelt2022,
author = {Jesper M. Skottfelt and Zoe Lee-Payne and David J. Hall and Ben Dryer},
title = {{Comparison of trap pumping and EPER data}},
volume = {12191},
booktitle = {X-Ray, Optical, and Infrared Detectors for Astronomy X},
editor = {Andrew D. Holland and James Beletic},
organization = {International Society for Optics and Photonics},
publisher = {SPIE},
pages = {121910A},
year = {2022},
doi = {10.1117/12.2627483},
URL = {https://doi.org/10.1117/12.2627483},
series = {}
}

@ARTICLE{Parsons2021,
       author = {{Parsons}, S. and {Buggey}, T. and {Holland}, A. and {Sembay}, S. and {Randall}, G. and {Hetherington}, O. and {Yeoman}, D. and {Hall}, D. and {Verhoeve}, P. and {Soman}, M.},
        title = "{Effects of temperature anneal cycling on a cryogenically proton irradiated CCD}",
      journal = {Journal of Instrumentation},
         year = 2021,
        month = nov,
       volume = {16},
       number = {11},
          eid = {P11005},
        pages = {P11005},
          doi = {10.1088/1748-0221/16/11/P11005},
archivePrefix = {arXiv},
       eprint = {2110.09081},
 primaryClass = {physics.ins-det},
       adsurl = {https://ui.adsabs.harvard.edu/abs/2021JInst..16P1005P}
}

@INPROCEEDINGS{C3TM,
       author = {{Skottfelt}, Jesper and {Hall}, David J. and {Dryer}, Ben and {Burgon}, Ross and {Holland}, Andrew},
        title = "{C3TM: CEI CCD charge transfer model for radiation damage analysis and testing}",
    booktitle = {High Energy, Optical, and Infrared Detectors for Astronomy VIII},
         year = 2018,
       editor = {{Holland}, Andrew D. and {Beletic}, James},
       series = {Society of Photo-Optical Instrumentation Engineers (SPIE) Conference Series},
       volume = {10709},
        month = jul,
          eid = {1070918},
        pages = {1070918},
          doi = {10.1117/12.2309944},
       adsurl = {https://ui.adsabs.harvard.edu/abs/2018SPIE10709E..18S}
}

@INPROCEEDINGS{Tarle2003,
       author = {{Tarle}, Gregory and {Akerlof}, Carl W. and {Aldering}, Greg and {Amanullah}, R. and {Astier}, Pierre and {Barrelet}, E. and {Bebek}, Christopher and {Bergstrom}, Lars and {Bercovitz}, John and {Bernstein}, Gary M. and {Bester}, Manfred and {Bonissent}, Alain and {Bower}, C.~R. and {Brown}, Mark L. and {Carithers}, William C., Jr. and {Commins}, Eugene D. and {Day}, C. and {Deustua}, Susana E. and {DiGennaro}, Richard S. and {Ealet}, Anne and {Ellis}, Richard S. and {Eriksson}, Mikael and {Fruchter}, Andrew and {Genat}, Jean-Francois and {Goldhaber}, Gerson and {Goobar}, Ariel and {Groom}, Donald E. and {Harris}, Stewart E. and {Harvey}, Peter R. and {Heetderks}, Henry D. and {Holland}, Steven E. and {Huterer}, Dragan and {Karcher}, Armin and {Kim}, Alex G. and {Kolbe}, William F. and {Krieger}, B. and {Lafever}, R. and {Lamoureux}, J. and {Lampton}, Michael L. and {Levi}, Michael E. and {Levin}, Daniel S. and {Linder}, Eric V. and {Loken}, Stewart C. and {Malina}, Roger and {Massey}, R. and {Miquel}, Ramon and {McKay}, Timothy and {McKee}, Shawn P. and {Moertsell}, E. and {Mostek}, N. and {Mufson}, Stuart and {Musser}, J.~A. and {Nugent}, Peter E. and {Oluseyi}, Hakeem M. and {Pain}, Reynald and {Palaio}, Nicholas P. and {Pankow}, David H. and {Perlmutter}, Saul and {Pratt}, R. and {Prieto}, Eric and {Refregier}, Alexandre and {Rhodes}, Jason and {Robinson}, Kem E. and {Roe}, N. and {Schubnell}, Michael S. and {Sholl}, Michael and {Smadja}, G. and {Smoot}, George F. and {Spadafora}, Anthony and {Tomasch}, Andrew D. and {Vincent}, D. and {von der Lippe}, H. and {Walder}, J. -P. and {Wang}, Guobin},
        title = "{SNAP NIR detectors}",
    booktitle = {IR Space Telescopes and Instruments},
         year = 2003,
       editor = {{Mather}, John C.},
       series = {Society of Photo-Optical Instrumentation Engineers (SPIE) Conference Series},
       volume = {4850},
        month = mar,
        pages = {919-926},
          doi = {10.1117/12.461774},
       adsurl = {https://ui.adsabs.harvard.edu/abs/2003SPIE.4850..919T}
}

@article{Hands2018,
author = {Hands, A. D. P. and Ryden, K. A. and Meredith, N. P. and Glauert, S. A. and Horne, R. B.},
title = {Radiation Effects on Satellites During Extreme Space Weather Events},
journal = {Space Weather},
volume = {16},
number = {9},
pages = {1216-1226},
doi = {https://doi.org/10.1029/2018SW001913},
url = {https://agupubs.onlinelibrary.wiley.com/doi/abs/10.1029/2018SW001913},
eprint = {https://agupubs.onlinelibrary.wiley.com/doi/pdf/10.1029/2018SW001913},
year = {2018}
}

@ARTICLE{Koldobskiy2022,
       author = {{Koldobskiy}, Sergey A. and {K{\"a}hk{\"o}nen}, Riikka and {Hofer}, Bernhard and {Krivova}, Natalie A. and {Kovaltsov}, Gennady A. and {Usoskin}, Ilya G.},
        title = "{Time Lag Between Cosmic-Ray and Solar Variability: Sunspot Numbers and Open Solar Magnetic Flux}",
      journal = {\solphys},
         year = 2022,
        month = mar,
       volume = {297},
       number = {3},
          eid = {38},
        pages = {38},
          doi = {10.1007/s11207-022-01970-1},
       adsurl = {https://ui.adsabs.harvard.edu/abs/2022SoPh..297...38K}
}

@article{Tahtinen2024,
	author = {T\"ahtinen, I. and Asikainen, Timo. and Mursula, K.},
	title = {Straight outta photosphere: Open solar flux without coronal modeling},
	DOI= "10.1051/0004-6361/202451267",
	url= "https://doi.org/10.1051/0004-6361/202451267",
	journal = {A&A},
	year = 2024,
	volume = 688,
	pages = "L32",
}

@article{Matthia2023,
title = {Active radiation measurements over one solar cycle with two DOSTEL instruments in the Columbus laboratory of the International Space Station},
journal = {Life Sciences in Space Research},
volume = {39},
pages = {14-25},
year = {2023},
note = {Radiation in human space exploration: Detectors and measurements, today and tomorrow},
issn = {2214-5524},
doi = {https://doi.org/10.1016/j.lssr.2023.04.002},
url = {https://www.sciencedirect.com/science/article/pii/S2214552423000299},
author = {Daniel Matthi\"a and Sanke Burmeister and Bartos Przybyla and Thomas Berger}
}

@ARTICLE{epaawol,
       author = {{Vehtari}, Aki and {Gelman}, Andrew and {Sivula}, Tuomas and {Jyl{\"a}nki}, Pasi and {Tran}, Dustin and {Sahai}, Swupnil and {Blomstedt}, Paul and {Cunningham}, John P. and {Schiminovich}, David and {Robert}, Christian},
        title = "{Expectation propagation as a way of life: A framework for Bayesian inference on partitioned data}",
      journal = {arXiv e-prints},
         year = 2014,
        month = dec,
          eid = {arXiv:1412.4869},
        pages = {arXiv:1412.4869},
          doi = {10.48550/arXiv.1412.4869},
archivePrefix = {arXiv},
       eprint = {1412.4869},
 primaryClass = {stat.CO},
       adsurl = {https://ui.adsabs.harvard.edu/abs/2014arXiv1412.4869V}
}

@MISC{2025hst..prop17925B,
       author = {{Bennet}, Paul and {Fardal}, Mark and {Kallivayalil}, Nitya and {McKinnon}, Kevin Andrew and {Patel}, Ekta and {Pawlowski}, Marcel and {Sohn}, Sangmo Tony and {Warfield}, Jack Thomas and {Watkins}, Laura L. and {van der Marel}, Roeland P.},
        title = "{20 Years of time baseline, The 3D kinematics of Centaurus A}",
 howpublished = {HST Proposal. Cycle 32, ID. \#17925},
         year = 2025,
        month = jan,
        pages = {17925},
       adsurl = {https://ui.adsabs.harvard.edu/abs/2025hst..prop17925B}
}

@ARTICLE{Mitra2021,
       author = {{Mitra}, Ayan and {Linder}, Eric V.},
        title = "{Cosmology requirements on supernova photometric redshift systematics for the Rubin LSST and Roman Space Telescope}",
      journal = {\prd},
         year = 2021,
        month = jan,
       volume = {103},
       number = {2},
          eid = {023524},
        pages = {023524},
          doi = {10.1103/PhysRevD.103.023524},
archivePrefix = {arXiv},
       eprint = {2011.08206},
 primaryClass = {astro-ph.CO},
       adsurl = {https://ui.adsabs.harvard.edu/abs/2021PhRvD.103b3524M}
}

@article{Soto2023,
    author = {Soto, Mario and Kuijken, Konrad and Rich, R Michael and Clarkson, William I and Nilo Castellón, José Luis and Fernández-Trincado, José G and Contreras Ramos, Rodrigo and Kunder, Andrea and Baravalle, Laura D and Alonso, M Victoria and Simion, Iulia T and Johnson, Christian I and Vieira, Katherine},
    title = {HST proper motions on the far side of the Galactic bar—data},
    journal = {Monthly Notices of the Royal Astronomical Society},
    volume = {524},
    number = {1},
    pages = {224-234},
    year = {2023},
    month = {06},
    issn = {0035-8711},
    doi = {10.1093/mnras/stad1911},
    url = {https://doi.org/10.1093/mnras/stad1911},
    eprint = {https://academic.oup.com/mnras/article-pdf/524/1/224/50798726/stad1911.pdf},
}

@ARTICLE{Ghez2008,
       author = {{Ghez}, A.~M. and {Salim}, S. and {Weinberg}, N.~N. and {Lu}, J.~R. and {Do}, T. and {Dunn}, J.~K. and {Matthews}, K. and {Morris}, M.~R. and {Yelda}, S. and {Becklin}, E.~E. and {Kremenek}, T. and {Milosavljevic}, M. and {Naiman}, J.},
        title = "{Measuring Distance and Properties of the Milky Way's Central Supermassive Black Hole with Stellar Orbits}",
      journal = {\apj},
         year = 2008,
        month = dec,
       volume = {689},
       number = {2},
        pages = {1044-1062},
          doi = {10.1086/592738},
archivePrefix = {arXiv},
       eprint = {0808.2870},
 primaryClass = {astro-ph},
       adsurl = {https://ui.adsabs.harvard.edu/abs/2008ApJ...689.1044G}
}

@INPROCEEDINGS{Holland1990,
       author = {{Holland}, A. and {Abbey}, A. and {Lumb}, D. and {McCarthy}, K.},
        title = "{Proton damage effects in EEV charge coupled devices.}",
    booktitle = {EUV, X-ray, and Gamma-ray instrumentation for astronomy},
         year = 1990,
       editor = {{Hudson}, Hugh S. and {Siegmund}, Oswald H.},
       series = {Society of Photo-Optical Instrumentation Engineers (SPIE) Conference Series},
       volume = {1344},
        month = nov,
        pages = {378-395},
          doi = {10.1117/12.23266},
       adsurl = {https://ui.adsabs.harvard.edu/abs/1990SPIE.1344..378H}
}

@incollection{Kilifarska2020,
title = {Chapter 5 - Galactic cosmic rays and solar particles in Earth's atmosphere},
editor = {Natalya A. Kilifarska and Volodymyr G. Bakhmutov and Galyna V. Melnyk},
booktitle = {The Hidden Link between Earth's Magnetic Field and Climate},
publisher = {Elsevier},
pages = {101-131},
year = {2020},
isbn = {978-0-12-819346-4},
doi = {https://doi.org/10.1016/B978-0-12-819346-4.00005-X},
url = {https://www.sciencedirect.com/science/article/pii/B978012819346400005X},
author = {Natalya A. Kilifarska and Volodymyr G. Bakhmutov and Galyna V. Melnyk}
}

@ARTICLE{Rowell2021,
       author = {{Rowell}, N. and {Davidson}, M. and {Lindegren}, L. and {van Leeuwen}, F. and {Casta{\~n}eda}, J. and {Fabricius}, C. and {Bastian}, U. and {Hambly}, N.~C. and {Hern{\'a}ndez}, J. and {Bombrun}, A. and {Evans}, D.~W. and {De Angeli}, F. and {Riello}, M. and {Busonero}, D. and {Crowley}, C. and {Mora}, A. and {Lammers}, U. and {Gracia}, G. and {Portell}, J. and {Biermann}, M. and {Brown}, A.~G.~A.},
        title = "{Gaia Early Data Release 3. Modelling and calibration of Gaia's point and line spread functions}",
      journal = {\aap},
         year = 2021,
        month = may,
       volume = {649},
          eid = {A11},
        pages = {A11},
          doi = {10.1051/0004-6361/202039448},
archivePrefix = {arXiv},
       eprint = {2012.02069},
 primaryClass = {astro-ph.IM},
       adsurl = {https://ui.adsabs.harvard.edu/abs/2021A&A...649A..11R}
}

@INPROCEEDINGS{Janesick2000,
       author = {{Janesick}, James R. and {McCarthy}, James K. and {Pinter}, Jeff H. and {Dosuoglu}, Taner},
        title = "{High-speed scientific CCDs: substrate bounce}",
    booktitle = {Optical and IR Telescope Instrumentation and Detectors},
         year = 2000,
       editor = {{Iye}, Masanori and {Moorwood}, Alan F.},
       series = {Society of Photo-Optical Instrumentation Engineers (SPIE) Conference Series},
       volume = {4008},
        month = aug,
        pages = {375-388},
          doi = {10.1117/12.395496},
       adsurl = {https://ui.adsabs.harvard.edu/abs/2000SPIE.4008..375J}
}

@ARTICLE{Bertin1996,
       author = {{Bertin}, E. and {Arnouts}, S.},
        title = "{SExtractor: Software for source extraction.}",
      journal = {\aaps},
         year = 1996,
        month = jun,
       volume = {117},
        pages = {393-404},
          doi = {10.1051/aas:1996164},
       adsurl = {https://ui.adsabs.harvard.edu/abs/1996A&AS..117..393B}
}

@inproceedings{ARRAKIHS,
      title = {ARRAKIHS: ESA’s new fast-implementation science mission},
      author = {{Corral van Damme}, C. and {Prod'Homme}, T. and {Isaak}, K. and {R{\"u}hl}, T. and {Sirianni}, M.},
      volume = {13092},
      booktitle = {Space Telescopes and Instrumentation 2024: Optical, Infrared, and Millimeter Wave},
      editor = {Laura E. Coyle and Shuji Matsuura and Marshall D. Perrin},
      organization = {International Society for Optics and Photonics},
      publisher = {SPIE},
      series = {},
      pages = {130920Q},
      year = {2024},
      doi = {10.1117/12.3020186},
      URL = {https://doi.org/10.1117/12.3020186}
}

@misc{plato2025,
      title={Impact of Charge Transfer Inefficiency on transit light-curves: A correction strategy for PLATO}, 
      author={Shaunak Mishra and Reza Samadi and Diane Bérard},
      year={2025},
      eprint={2510.22092},
      archivePrefix={arXiv},
      primaryClass={astro-ph.EP},
      url={https://arxiv.org/abs/2510.22092}, 
}

@inproceedings{Clarke2025,
      title = {Device modelling and model verification for the Euclid CCD273 detector},
      author = {A. Clarke and D. Hall and N. Murray and A. Holland and D. Burt},
      volume = {8453},
      booktitle = {High Energy, Optical, and Infrared Detectors for Astronomy V},
      editor = {Andrew D. Holland and James W. Beletic},
      organization = {International Society for Optics and Photonics},
      publisher = {SPIE},
      series = {},
      pages = {84531I},
      year = {2012},
      doi = {10.1117/12.925887},
      URL = {https://doi.org/10.1117/12.925887}
}

\bsp	
\label{lastpage}

\end{document}